 \documentclass[acmsmall,screen]{acmart}
\usepackage{microtype}
\UseMicrotypeSet[protrusion]{basicmath} % 可选
\AtBeginDocument{%
  } 
\setcopyright{acmlicensed}
\copyrightyear{2018}
\acmYear{2018}
\acmDOI{XXXXXXX.XXXXXXX}

\acmJournal{JACM}
\acmVolume{37}
\acmNumber{4}
\acmArticle{111}
\acmMonth{8}

\usepackage[pagewise]{lineno}
\usepackage{graphicx}
\usepackage{subcaption} 
\usepackage{amsmath}
\usepackage{amsthm}

\usepackage{ragged2e}
\usepackage{tabularx}  
\usepackage{bm}
\usepackage[normalem]{ulem} % 不影响 \emph

\usepackage{listings,framed,xcolor}
\definecolor{lgray}{rgb}{0.95,0.95,0.95}
\usepackage[table]{xcolor}

\usepackage{float}
\usepackage{booktabs}
\usepackage{multirow}
\usepackage{algorithmic}
\usepackage[ruled,linesnumbered]{algorithm2e}
\SetKwRepeat{Do}{do}{while}

\usepackage{enumitem}

\newlist{rqsublist}{itemize}{1}
\setlist[rqsublist]{label={}, leftmargin=1.8em, itemsep=0pt, topsep=0pt}

\usepackage{rotating}
\usepackage{balance}
\usepackage{threeparttable}
\usepackage{makecell}
\usepackage{pdflscape}

\usepackage{xstring}
\usepackage{seqsplit}

\usepackage{graphicx}

\usepackage[export]{adjustbox}

\usepackage{stfloats}
\usepackage{verbatim}
\usepackage{setspace}
\usepackage{threeparttable}
\usepackage{supertabular,booktabs}
\usepackage{rotating}
\usepackage{longtable}
\usepackage{supertabular}
\usepackage{pifont}
\usepackage{xcolor,pict2e}% to allow any radius

\usepackage{lettrine}
\usepackage{hyperref}
\hypersetup{
    colorlinks=true,
    linkcolor=blue,    % 超链接颜色为蓝色
    citecolor=blue,    % 引用链接颜色为蓝色
    urlcolor=black      % URL 链接颜色为蓝色
}
\usepackage{xurl}      % 核心：支持 pdflatex 下的 URL 自动换行

\makeatletter
\renewcommand\@cite[2]{%
    {\color{blue}[#1]}% 修正此处
}
\makeatother
\usepackage{fancyhdr}

\allowdisplaybreaks[4] %公式跨页

\usepackage{tcolorbox}
\tcbuselibrary{breakable} %跨页盒子
\usepackage{xcolor}
\usepackage{soul}   % for highlighting

\newcounter{ChenhuiCommentCounter}
\definecolor{lightpink}{rgb}{1.0, 0.87, 0.87}
\definecolor{lightpurple}{rgb}{0.94, 0.85, 0.94}
\definecolor{lightyellow}{rgb}{1.0, 1.0, 0.88}
\definecolor{lightorange}{rgb}{1.0, 0.882, 0.788}
\definecolor{lightblue}{RGB}{204, 224, 238}
\definecolor{lightgreen}{rgb}{0.82, 0.97, 0.82}
\definecolor{darkorange}{rgb}{0.92, 0.60, 0.38}   % deeper warm orange
\definecolor{darkblue}{RGB}{102, 153, 204}        % deeper steel/cornflower blue
\definecolor{darkgreen}{rgb}{0.45, 0.75, 0.45}    % deeper soft green

\definecolor{darkorange2}{rgb}{0.85, 0.45, 0.20}
\definecolor{darkblue2}{RGB}{70, 120, 180}
\definecolor{darkgreen2}{rgb}{0.25, 0.60, 0.25}

\newcommand{\builder}{\textsc{AtgBuilder}}
\newcommand{\fastbot}{\textsc{FastBot2}}
\newcommand{\monkey}{\textsc{Monkey}}
\newcommand{\humanoid}{\textsc{Humanoid}}
\newcommand{\qtest}{\textsc{Q-Testing}}
\newcommand{\scenedroid}{\textsc{SceneDroid}}
\newcommand{\stoat}{\textsc{Stoat}}
\newcommand{\ape}{\textsc{APE}}
\newcommand{\frontmatter}{\textsc{Frontmatter}}
\newcommand{\archidroid}{\textsc{ArchiDroid}}

\DeclareRobustCommand{\revp}[1]{{\color{black}#1}}

\newcommand{\hlo}[1]{%
  \begingroup
  \setlength{\fboxsep}{0pt}%
  \colorbox{lightorange}{\strut #1}%
  \endgroup
}

\newcommand{\hlb}[1]{%
  \begingroup
  \setlength{\fboxsep}{0pt}%
  \colorbox{lightblue}{\strut #1}%
  \endgroup
}

\newcommand{\hlg}[1]{%
  \begingroup
  \setlength{\fboxsep}{0pt}%
  \colorbox{lightgreen}{\strut #1}%
  \endgroup
}
\newcommand{\textttsplie}[1]{\texttt{\seqsplit{#1}}}

\usepackage{xcolor}
\usepackage{tikz}
\usetikzlibrary{decorations.pathmorphing}

\newcommand{\rwavo}[1]{%
  \tikz[baseline=(X.base)]{
    \node[inner sep=0pt, outer sep=0pt] (X) {$#1$};
    \draw[darkorange2, line width=0.75pt,
          decorate, decoration={snake, amplitude=0.15mm, segment length=1.3mm}]
      ([yshift=-0.3mm]X.south west) -- ([yshift=-0.3mm]X.south east);
  }%
}

\newcommand{\rwavb}[1]{%
  \tikz[baseline=(X.base)]{
    \node[inner sep=0pt, outer sep=0pt] (X) {$#1$};
    \draw[darkblue2, line width=0.75pt,
          decorate, decoration={snake, amplitude=0.15mm, segment length=1.3mm}]
      ([yshift=-0.3mm]X.south west) -- ([yshift=-0.3mm]X.south east);
  }%
}

\newcommand{\rwavg}[1]{%
  \tikz[baseline=(X.base)]{
    \node[inner sep=0pt, outer sep=0pt] (X) {$#1$};
    \draw[darkgreen2, line width=0.75pt,
          decorate, decoration={snake, amplitude=0.15mm, segment length=1.3mm}]
      ([yshift=-0.3mm]X.south west) -- ([yshift=-0.3mm]X.south east);
  }%
}

\newcommand{\rqsummary}[2]{%
\begin{tcolorbox}[
  breakable,
  colframe=orange,
  arc=1mm,
  left=1mm, right=1mm, top=0mm, bottom=0mm,
  boxrule=0.25mm,
  before skip=2pt,  
  after skip=0pt, 
  before upper={\raisebox{-0.2\height}{\includegraphics[scale=0.065]{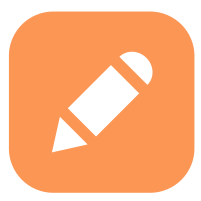}}~}
]
\textbf{Summary of Answers to RQ#1:}
#2
\end{tcolorbox}%
}
 
\newcounter{DaveCommentCounter}
\newcommand{\ddu}[1]{
    \stepcounter{DaveCommentCounter}
    \textcolor{blue}{\textit{/**Dave's comment [\arabic{DaveCommentCounter}]: I don't understand the intended meaning in the next sentence. Please revise/delete/explain. **/}}
}

\newcommand{\dns}[1]{
    \stepcounter{DaveCommentCounter}
    \textcolor{blue}{\textit{/**Dave's comment [\arabic{DaveCommentCounter}]: I'm not sure that I have captured the intended meaning in the next sentence. Please check/confirm. **/}}
}

\newcounter{AbelCommentCounter}
\newcounter{RubingCommentCounter}
\begin{document}
\title{\builder: Feature-Assisted Graph Learning for Activity Transition Graph Construction with Seed Supervision}

\author{Chenhui Cui}
\email{3230002105@student.must.edu.mo}
\orcid{0009-0004-8746-316X}
\affiliation{
  \institution{School of Computer Science and Engineering, Macau University of Science and Technology}
  \city{Macau SAR}
  \postcode{999078}
  \country{China}
}
\author{Zixiang Xian}
\email{3220001352@student.must.edu.mo}
\orcid{000-0002-8892-6187}
\affiliation{
  \institution{School of Computer Science and Engineering, Macau University of Science and Technology}
  \city{Macau SAR}
  \postcode{999078}
  \country{China}
}
\author{Danyu Li}
\email{3230002471@student.must.edu.mo}
\orcid{0009-0001-8884-152X}
\affiliation{
  \institution{School of Computer Science and Engineering, Macau University of Science and Technology}
  \city{Macau SAR}
  \postcode{999078}
  \country{China}
}

\author{Tao Li}
\email{3220007015@student.must.edu.mo}
\orcid{0009-0001-7413-9692}
\affiliation{
  \institution{School of Computer Science and Engineering, Macau University of Science and Technology}
  \city{Macau SAR}
  \postcode{999078}
  \country{China}
}

\author{Rubing Huang}
\email{rbhuang@must.edu.mo}
\orcid{0000-0002-1769-6126}
\affiliation{
  \institution{School of Computer Science and Engineering, Macau University of Science and Technology}
  \city{Macau SAR}
  \postcode{999078}
  \country{China}  
}
  \affiliation{
  \institution{Macau University of Science and Technology Zhuhai MUST Science and Technology Research Institute}
  \city{Zhuhai}
  \postcode{519099}
  \country{China}
}

\author{Dave Towey}
\email{dave.towey@nottingham.edu.cn}
\orcid{0000-0003-0877-4353}
\affiliation{
  \institution{School of Computer Science, University of Nottingham Ningbo China}
  \city{Ningbo}
  \postcode{315100}
  \country{China}
}

\author{Shikai Guo}
\email{shikai.guo@dlmu.edu.cn}
\orcid{0000-0002-8554-6365}
\affiliation{
  \institution{School of Information Science and Technology, Dalian Maritime University}
  \city{Dalian}
  \postcode{116026}
  \country{China}   
}

\author{Jiakun Liu}
\email{jiakunliu@hit.edu.cn}
\orcid{0000-0002-7273-6709}
\affiliation{
  \institution{Faculty of Computing, Harbin Institute of Technology}
  \city{Harbin}
  \postcode{150001}
  \country{China}
  }

\begin{abstract}
    Android applications (apps) have become indispensable in daily life.
    They can be organized around activities that provide visual \textit{Graphical User Interface} (GUI) containers that host the UI and handle user interaction events.
    To help better understand their behavior, \textit{Activity Transition Graphs} (ATGs) have been widely used to model apps' GUI navigation. 
    However, the construction of high-quality ATGs is challenging:
    ATGs based on static analysis may miss acceptable transitions and may extract infeasible ones;
    while dynamically explored ATGs can yield incomplete transitions.
    Recent learning-based approaches can treat ATG construction as a seed-supervised link-prediction task.
    However, the use of activity-layout and widget-trigger information for ATG construction remains limited.  
    To address this, we propose \builder, a feature-assisted graph-learning approach for seed-supervised ATG construction.
    \builder\ uses a \textit{Large Language Model} (LLM) to summarize UI activity metadata from layouts into compact textual functionality summaries:
    These summaries can then be encoded using a sentence-embedding model.
    \builder\ explicitly models widget-trigger information into the edge attribute (rather than mixing it into activity representations):
    It then uses an auxiliary widget-attribute reconstruction objective on this information during model training.
    \builder's performance was evaluated across a series of ablations on the \frontmatter\ corpus, and an experiment on a 98-app benchmark using manually-checked ground-truth ATGs.  
    The results show that \builder\ outperformed the state-of-the-art (SOTA) methods in constructing effective ATGs:
    On the 98-app benchmark, \builder\ improved the F1-score by 15.41\% to 77.34\% over SOTA baselines. 
    \revp{We further evaluated its practical usefulness by using \builder-predicted ATGs as lightweight navigation guidance for \monkey, \ape, and \fastbot{}
    ---
    three automated GUI-exploration tools.
    This improved the activity and transition coverage, suggesting the potential to also improve existing automated GUI-exploration tools.}    
\end{abstract}




\received{25 June 2026}

\maketitle

\section{Introduction} 
\label{sec:introduction}

In the third quarter of 2025, Android maintained its leading position in the global mobile operating system market, with approximately 73.75\% of the market share~\cite{android202511}.
Android applications (apps) span a broad spectrum, from e-commerce and navigation tools to fitness and creative design platforms~\cite{googleplay_categories}.
These apps have become indispensable tools in daily life.

Android apps can be organized around \emph{activities}~\cite{mobiletestingsurvey2018}:
These serve as \textit{Graphical User Interface} (GUI) containers for rendering UI pages and handling user-interaction events.
To understand the apps' functionality, \textit{Activity Transition Graphs} (ATGs) have been used to model feasible activity-to-activity navigation as directed graphs~\cite{storydroid2019,fazzinipd2018,zhaoyixue2018,iconintent2019,archidroid2023,atgemprical2025}.
They have been widely used to capture UI-page reachability in mobile apps, especially in the context of automated GUI testing~\cite{storydroid2019,fazzinipd2018,zhaoyixue2018,iconintent2019}.
\citet{julia_rubin_goal_explorer}, for example, used ATGs to guide GUI testing, delivering higher coverage within limited budgets.
\revp{ATGs provide a structural model for reasoning about activity reachability, and for planning activity-transition exploration~\cite{atgemprical2025}. 
Compared with finer-grained graph representations (such as \textit{UI-page Transition Graphs} (UTGs)~\cite{tacdroid2025})
---
which usually distinguish pages, UI states, and UI components within the same activity
---
activity-level granularity provides a more fundamental navigation layer.
If these transitions cannot be modeled accurately, then the more fine-grained elements are even more challenging.}

Traditional ATG-construction methods can be broadly categorized into either \textit{static analysis} or \textit{dynamic exploration}.
Static-analysis approaches~\cite {staticwtg2018,KuznetsovAGZ18,frontmatter2021,permdroid2022} extract GUI-related metadata and transitions directly from APKs
, constructing the ATGs without app execution.
This enables the creation of large-scale ATG corpora across many apps.
\frontmatter~\cite{frontmatter2021}, for example, uses this approach for ATG construction.
However, \textbf{statically-analyzed ATGs can introduce both \textit{false-negative}  and \textit{false-positive} transitions 
---
missing feasible transitions, and including infeasible transitions at runtime.}
This can reduce the testing efficacy:
Missing feasible transitions could remove the reachability of certain activities, and including infeasible transitions at runtime could waste computing resources.
Dynamic-exploration approaches~\cite{monkey,stoat2017,ape2019,qtesting2020,humanoid2019,fastbot22022,scenedroid2023}, in contrast, collect and refine transitions during execution~\cite{atgemprical2025}.
These ATGs are generally more trustworthy than those derived purely from static analysis~\cite{timemachine2020,fastbot22022}.
However, because of limited budgets, \textbf{dynamic exploration can yield precise, but incomplete, transitions}:
The observed transitions are trustworthy, but many feasible transitions are never exercised.

Recent studies have explored ATG construction using learning-based approaches that combine deep learning with static analysis or dynamic exploration.
These approaches~\cite{archidroid2023,tacdroid2025} view ATG construction (and augmentation) as a seed-supervised link-prediction task~\cite{linkprediction2018} that can be addressed using \textit{Graph Neural Networks} (GNNs)~\cite{gnnsurvey2021}.
\archidroid~\cite{archidroid2023}, for example, augments a seed ATG using a \textit{Graph Convolutional Network} (GCN) that jointly encodes activity identifiers (activity names) and the ATG structure to predict missing transitions.
\textsc{TacDroid}~\cite{tacdroid2025} builds \textit{UI-page Transition Graphs} (UTGs) through hybrid static and dynamic analyses: 
Although this is not ATG construction, it highlights the benefit of modeling transitions with richer UI-page information.
\textbf{These previous studies, however, have not taken advantage of activity-layout and widget-trigger information for transition modeling.}
(1) Because activities use rich UI layouts to implement their functionality, and similar functionality could be realized with very different layouts, it may be difficult to identify functional similarity directly from raw layout structures.
(2) Activity transitions are often triggered by specific widgets, but previous learning-based ATG-construction methods have typically only modeled transitions as activity pairs.
When transitions are modeled only as activity pairs, the triggering widget is not represented:
This can result in different widget-triggered navigations between the same activity pair becoming indistinguishable.

Based on these observations, in this paper, we propose \builder, a feature-assisted graph-learning approach for seed-supervised ATG construction.
\builder\ augments analysis-derived seed ATGs with static UI metadata extracted from APKs using \frontmatter~\cite{frontmatter2021}. 
The extracted metadata includes per-activity identifiers, UI layouts, and widget-trigger information.
\textit{To address the limitation of activity representation}, \builder\ does not rely only on raw UI layout structures as the model inputs:
It uses a \textit{Large Language Model} (LLM)~\cite{summary2025} to summarize layout-related UI metadata into a compact functional description for each activity, encoding the summaries using a sentence-embedding model.
This semantic representation emphasizes activity functionality over implementation-specific layout details for transition modeling.
\builder\ explicitly models widget-trigger information as activity-independent edge features (rather than mixing it into activity representations).
Furthermore, \builder\ uses an auxiliary widget-attribute to preserve widget-trigger information, while focusing on transition-existence prediction as the primary task.
To evaluate the performance and usefulness of \builder, we 
(1) conducted ablation studies on the \frontmatter\ corpus~\cite{frontmatter2021}; 
(2) compared \builder\ with seven state-of-the-art (SOTA) baselines on a 98-app benchmark set with manually-checked ground-truth ATGs~\cite{atgemprical2025}; and 
\revp{(3) evaluated whether or not \builder-predicted ATGs could guide automated GUI-exploration tools.}
The results show that \builder\ can improve on the SOTA F1-score by between 15.41\% and 77.34\%.
\revp{Furthermore, \builder-predicted ATGs used as lightweight GUI-navigation guidance for \monkey~\cite{monkey}, \ape~\cite{ape2019}, and \fastbot~\cite{fastbot22022}
improved both the activity and transition coverage:
This highlights the potential for \builder\ to guide GUI exploration.}

The main contributions of this paper are as follows:
\begin{itemize}[leftmargin=2em, topsep=0pt, itemsep=0pt, parsep=0pt, partopsep=0pt]
    \item 
    We propose \builder, a feature-assisted graph-learning approach for seed-supervised ATG construction that augments analysis-derived seed ATGs with static UI metadata, and predicts missing transitions.
    
    \item 
    We propose a functionality-oriented activity representation based on layout-related UI metadata (summarized by an LLM).

    \item 
	We incorporate widget-trigger information in the transition modeling, and propose an auxiliary objective to reconstruct widget attributes during training (thereby preserving widget information).
	    
	\item
    \revp{We report on a comprehensive evaluation of model training under different ablations, a 98-app ground-truth benchmark with seven SOTA baselines, and an ATG-guided GUI-exploration study over automated tools.}
\end{itemize}

The rest of this paper is organized as follows:
Section~\ref{sec:motivation} presents motivating examples.
Section~\ref{sec:method} presents the \builder\ design.
Sections \ref{sec:experimental-study} and \ref{sec:experimental-results} report the experimental setup and results.
Section~\ref{sec:discussions} discusses the key findings and implications.
Finally, Section~\ref{sec:conclusions} concludes the paper and outlines some potential future work.

\section{Motivating Examples}
\label{sec:motivation}

\begin{figure}
    \centering
    \includegraphics[width=\linewidth]{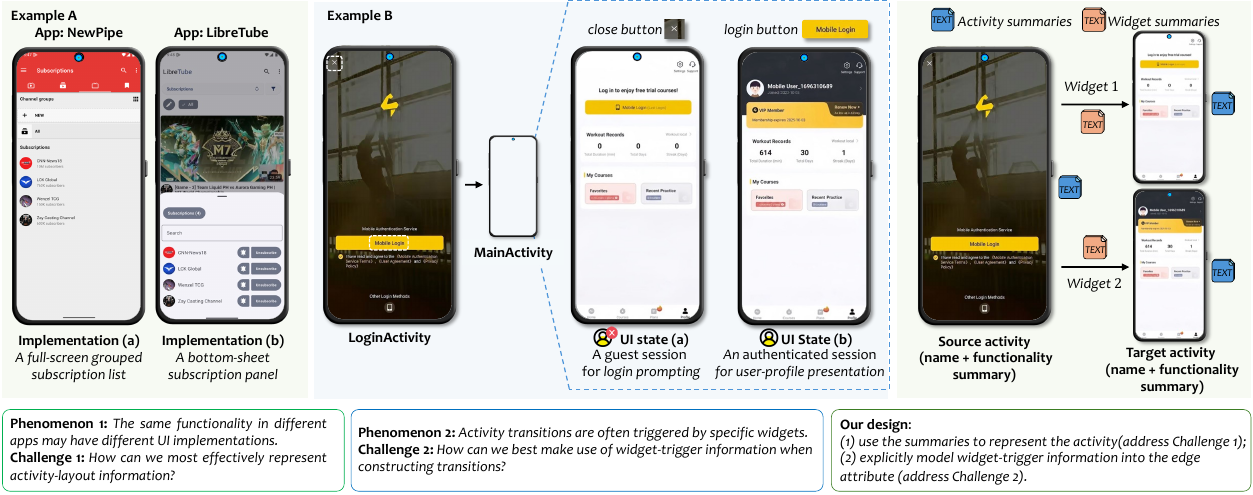}
    \caption{An illustrative example of our motivation and design.}
    \label{fig:motivation}
\end{figure}

This section provides some examples that motivate our study.
Example~A in Figure~\ref{fig:motivation} shows the subscription pages for two open-source YouTube clients: 
NewPipe~\cite{newpipe} and LibreTube~\cite{libretube}.
An observation is:
\textbf{The same functionality in different apps may have different UI implementations (Phenomenon 1).}
This leads to the following challenge when constructing ATGs:

\textbf{Challenge 1: How can we most effectively represent activity-layout information?}
A natural intuition is to encode activities using their UI layout as features~\cite{tacdroid2025}.
However, different UI implementations can vary significantly, even for the same functionality:
NewPipe~\cite{newpipe}, for example, presents the subscription page as a full-screen list, whereas LibreTube~\cite{libretube} uses a bottom-sheet subscription panel.
Compared with NewPipe, LibreTube adopts a multi-layered UI design (a bottom-sheet panel over the main feed/player view) and richer list-item widgets (e.g., notification and unsubscribe buttons).
Such implementation-specific variations make layout features less reliable when learning functionality-oriented activity representations, and can mislead the model into focusing on UI structure rather than functionality.

\textbf{To address this challenge, we use an LLM to summarize layout-related UI metadata in terms of functionality, using the summaries to represent the activity.}
Functionality is treated as the main activity characteristic, with concrete UI layouts being considered as merely implementation choices for the functionality.
LLMs have a strong ability to understand UI layout information~\cite{screenai2024,screenread2024}, and can also produce concise functional summaries~\cite{screenai2024,summary2025}. 
Our design, therefore, uses an LLM to map complex UI layouts into a normalized functionality representation.
This does not rely on raw UI layout structures as inputs, but rather uses LLM-generated functional descriptions to distinguish activities.
\revp{Similar functionality summaries do not determine the same transitions:
They provide normalized activity-level semantic features, with the final transition decision being learned from activity features, graph context, and widget-trigger information.}

Example~B in Figure~\ref{fig:motivation} shows a login page (\texttt{\seqsplit{LoginActivity}}) and a home page (\texttt{\seqsplit{MainActivity}}) from a fitness app, and leads to another observation:
\textbf{Activity transitions are often triggered by specific widgets (Phenomenon 2).}
Previous study~\cite{archidroid2023} typically represented navigation only with the source and destination activities, without including which widget triggered the transition.
This motivated us to model widget-trigger information, leading to the following challenge:

\textbf{Challenge 2: How can we best make use of widget-trigger information when constructing transitions?}
Although widgets appear in an activity's UI layout, an activity-pair transition cannot identify which specific widget triggers the transition.
Activities may often contain many registered widgets, obscuring the true widget-trigger information.
Example~B in Figure~\ref{fig:motivation}, for example, shows that the transition from \texttt{LoginActivity} to \texttt{MainActivity} can be triggered by different widgets (the \emph{close} button and the \emph{login} button):
With an activity-pair representation, the model may not easily distinguish which widget triggers the transition.
Furthermore, a simple concatenation of the widget-trigger features into the activity embeddings would introduce redundancy.

\textbf{To address this challenge, we explicitly model widget-trigger information into the edge attribute (rather than mixing it into activity representations).}
This edge-attribute modeling matches how the navigation is executed:
It avoids mixing up the true trigger with irrelevant widgets in the same activity, and helps distinguish transitions that share the same activity pair, but are triggered by different widgets.
Because the activity representation is based on an LLM-generated summary, which may lack specific trigger widget details,
the widget-trigger information should be modeled as an activity-independent edge attribute.
\textbf{We also introduce an additional widget-attribute reconstruction objective, to support model training:}
By reconstructing widget-trigger information on transition edges, the model is encouraged to learn widget features as well as activity representations, ensuring that the widget-trigger information is preserved.
However, transition prediction remains the primary task, with the reconstruction objective acting as an auxiliary regularizer:
The model uses widget-trigger information when it is available, but uses a default feature (to ensure stability) when it is missing.

\section{\builder}
\label{sec:method}

\revp{
To make effective use of activity-layout and widget-trigger information for transition modeling, we propose \builder\ for seed-supervised ATG construction, defined as follows:

\textsc{\textbf{Definition (Activity Transition Graph):}}
For an Android app, an Activity Transition Graph (ATG) is a directed graph $\mathcal{G}=(V,E)$, where 
$V=\{v_1,\ldots,v_n\}$ is a finite set of Android activities, and 
$E\subseteq \{(v_i,v_j)\mid v_i,v_j\in V, i\neq j\}$ is a set of directed activity-level transitions.
Each node $v_i\in V$ denotes one Android activity. 
Each directed edge $(v_i,v_j)\in E$ denotes a possible transition from source activity $v_i$ to a different destination activity $v_j$:
Self-loops are excluded from $E$, as they do not represent inter-activity transitions.}

\begin{figure}
    \centering
    \includegraphics[width=\linewidth]{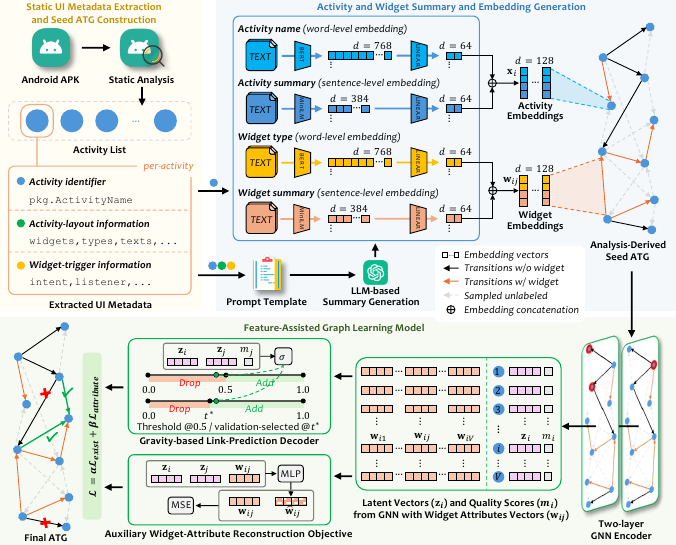}%\captionsetup{skip=2pt}
    \caption{\builder\ framework.}  
    \label{fig:framework}   
\end{figure}

Figure~\ref{fig:framework} presents an overview of the main \builder\ framework:
Given an APK, \builder\ 
\revp{(1) extracts static UI metadata, and constructs a seed ATG that provides activity nodes, transition candidates, and available widget-trigger information;
(2) transforms activity-level functionality and widget-trigger information into fixed-size node and edge attributes (using the summary and embedding-generation module); and
(3) trains a GNN-based link predictor on the activity-level graph with an auxiliary reconstruction objective that encourages the model to retain widget-trigger information during training.}

\subsection{Static UI Metadata Extraction and Seed ATG Construction}
\label{sec:method-static}

We first perform static analysis on the APK to extract UI metadata and to construct a seed ATG for supervised learning.
The extracted metadata serves two purposes:
(1) deriving seed transitions for supervision; and 
(2) providing activity-layout and widget-trigger information for feature construction during model training.
We use \frontmatter~\cite{frontmatter2021} to perform the static analysis to obtain three types of metadata:
\begin{itemize}[leftmargin=2em, topsep=0pt, itemsep=0pt, parsep=0pt, partopsep=0pt]
    \item 
    \textit{Activity identifiers} (activity names) uniquely index activities in an app, and have been used in previous work~\cite{archidroid2023,tacdroid2025,utg2025}.
    \builder\ uses the fully-qualified activity name (e.g., \textttsplie{com.app.MainActivity}) as the identifier.
    No naming constraints are imposed to maximize coverage across diverse naming conventions.
    For example, \textsc{ArchiDroid}~\cite{archidroid2023} restricts identifiers to those ending with the conventional \textttsplie{*Activity} suffix, which excludes many real-world activities.
    The identifiers define the activity set for each app, and are encoded as activity-name embeddings for ATG-node-feature construction.

    \item 
    \textit{Activity-layout information} captures more detailed per-activity UI-layout information, beyond the identifiers.
    \frontmatter~\cite{frontmatter2021} is used to extract a UI-element tree from layout resources, for each activity.
    This tree records activity-layout information (such as the view class, resource/view ID, and visible text (when available)), for each UI element, along with parent-child relations for all UI elements.
    The layout tree is passed as contextual input to the LLM for summarization (Section~\ref{sec:method-embed}).

    \item 
    \textit{Widget-trigger information} is the key metadata that enables trigger-aware transition modeling.
    \builder\ uses \frontmatter~\cite{frontmatter2021} to statically extract widget-event signals from the app logic, and align them to the UI widgets.
    The widget-trigger information (including the widget-UI structure and widget-event signals) is used to generate widget summaries (Section~\ref{sec:method-embed}) and construct ATG-edge features.
\end{itemize}

An analysis-derived seed ATG can then be constructed, where nodes are activities and edges are analysis-derived transitions.
\revp{Each activity node can carry node attributes derived from its identifier and functionality summary; and 
each candidate directed edge can carry edge attributes derived from widget-trigger information.}

\subsection{Activity and Widget Summary and Embedding Generation}
\label{sec:method-embed}

At this point, we have obtained an analysis-derived seed ATG with activity-layout and widget-trigger information.
However, as discussed in Section~\ref{sec:motivation}, UI metadata can be implementation-specific across apps:
The extracted metadata is therefore converted into compact functionality-oriented summaries, and both identifiers and summaries are embedded into fixed-size vectors for feature fusion and transition prediction.

\subsubsection{LLM-based Summary Generation}
To avoid implementation-specific layout details across activities and widgets, \builder\ uses an LLM to generate concise functionality summaries.
GPT-4o~\cite{gpt4o} is prompted with the activity-layout and widget-trigger information for each activity and widget in the seed ATG, and asked to produce a one-sentence functional summary (up to 30 words)\footnote{\revp{Due to space limitations, the full prompt template, the preliminary study for determining the summary-length setting, additional comparison details, and reproducibility details are reported in the Appendix~\ref{sec:supp-prompt} to Appendix~\ref{sec:supp-gcn-gin}.}}.
This step converts the detailed UI metadata in the seed ATG into one-sentence summaries.

\subsubsection{Embedding Encoding and Feature Fusion}
To support transition prediction, \builder\ vectorizes activity identifiers, widget types, and the LLM-generated summaries into a unified feature representation:
(1) activity identifiers and widget types are encoded using a word-level encoder (e.g., BERT-based embeddings~\cite{archidroid2023,tacdroid2025}); and 
(2) activity/widget summaries are encoded using a sentence-embedding model (e.g., Sentence-BERT style embeddings~\cite{sentencebert2019}).
Because these encoders produce different dimensionalities, a shared linear projection is used to map each embedding to a common dimension, $d_{\mathrm{pro}}$:
$\mathbf{x}_{\mathrm{pro}}=\mathbf{x}\mathbf{W}$, where $\mathbf{W}\in\mathbb{R}^{d_{\mathrm{in}}\times d_{\mathrm{pro}}}$.
Following common practice in prior work~\cite{tacdroid2025}, $d_{\mathrm{pro}}$ is set to $64$.

Features are then fused by concatenating specific embeddings:
(1) for each activity node, its projected identifier embedding and its projected activity-summary embedding are concatenated; and
(2) for each transition edge, the projected widget-type embedding and the projected widget-summary embedding are concatenated.
This yields $2d_{\mathrm{pro}}$-dimensional fused vectors for activity-nodes and transition-edges ($128$ dimensions in our implementation).

Finally, each activity node and transition edge has been converted into a fixed-size fused feature vector, yielding a feature-embedded seed ATG for subsequent graph learning.

\subsection{Feature-Assisted Graph Learning Model}
\label{sec:method-model}

Following previous work~\cite{tacdroid2025,archidroid2023}, we consider ATG construction as a seed-supervised link-prediction problem:
The goal is to learn a link predictor that estimates the existence of activity-to-activity transitions for a given feature-embedded seed ATG.

For each app, a seed ATG $\mathcal{G}_{\mathrm{seed}}=(V, E)$ is defined, where each node $v\in V$ denotes an activity with node feature $\mathbf{x}_v$.
Each candidate edge $(i,j)\in E$ corresponds to an activity pair $v_i$ and $v_j$, and is associated with a binary label $y_{ij}\in\{0,1\}$:
$y_{ij}=1$ indicates a seed transition, and $y_{ij}=0$ indicates a non-transition.
Each edge is also associated with an edge-feature vector $\mathbf{w}_{ij}$, representing its widget-trigger information.
If a concrete trigger widget cannot be identified, then a predefined placeholder attribute (e.g., \textttsplie{NONE\_WIDGET}) is used to ensure that $\mathbf{w}_{ij}$ remains well-defined.
The model outputs a transition probability $\hat{p}_{ij}\in[0,1]$ for each $(i,j)\in E$.
The predicted ATG $\mathcal{G}_{\mathrm{pred}}$ is then constructed by retaining edges whose probabilities exceed a decision threshold.
The threshold selection and decision details are discussed in Section~\ref{sec:alpha_beta}.

The proposed graph-learning model consists of three components:
(1) a GNN encoder that computes activity representations;
(2) a gravity-based decoder that predicts transition existence; and
(3) an auxiliary widget-attribute reconstruction objective that regularizes training, preserving widget-trigger information.

\subsubsection{GNN Encoder}
A GNN encoder computes latent activity representations that integrate node features with graph connectivity:
These are then used to estimate the existence of transitions on candidate edges in $\mathcal{G}_{\mathrm{seed}}$.
A preliminary comparison (Appendix~\ref{sec:supp-gcn-gin}) of a \textit{Graph Isomorphism Network} (GIN)~\cite{gin2019} and a \emph{Graph Convolutional Network} (GCN) (used in previous work~\cite{archidroid2023,tacdroid2025}) revealed the GIN to have the better performance, so it became the default encoder in this study.
Given node features $\mathbf{x}_v$ and an edge set $E$, the encoder produces a latent vector $\mathbf{z}_v$ for each activity node.
The encoder uses a two-layer architecture with ReLU activations and dropout.

The encoder outputs a latent embedding matrix $\mathbf{Z}=[\mathbf{z}_1;\ldots;\mathbf{z}_{|V|}]$, based on which an \textit{attractiveness score} for each activity is parameterized:
\revp{$m_j=\mathbf{u}^\top \mathbf{z}_j + c$,}
\revp{where 
$\mathbf{z}_j$ is the $d_{\mathrm{pro}}$-dimensional real-valued latent embedding of destination activity $v_j$; and 
$\mathbf{u}$ and $c$ are the learned weight vector and scalar bias of the linear scoring layer, respectively.
The scoring parameters $\mathbf{u}$ and $c$ are jointly optimized, using the encoder and decoder, through backpropagation.}
The attractiveness score, \revp{$m_j\in\mathbb{R}$}, describes the tendency for activity $v_j$ to attract incoming transitions.

\subsubsection{Gravity-based Link-Prediction Decoder}
\label{sec:decoder}
Given the latent activity embeddings, $\mathbf{z}_i$ and $\mathbf{z}_j$ (produced by the GNN encoder), a gravity-based decoder~\cite{tacdroid2025} produces an existence probability for each candidate transition edge based on the activity features.

Intuitively speaking, a transition is more likely when the destination activity is more attractive and when the two activities are closer in the embedding space.
For a candidate edge $(i,j)\in E$, the squared Euclidean distance,
$d_{ij}=\|\mathbf{z}_i-\mathbf{z}_j\|_2^2$, is first computed:
This measures the dissimilarity between the source and destination activities.
The destination attractiveness score $m_j$ is then combined with the distance term to obtain the edge score:
$s_{ij}=m_j-\log(d_{ij}+\epsilon)$,
where $\epsilon$ is a small constant for numerical stability.
\revp{Finally, the decoder obtains the transition-existence probability as $\hat{p}_{ij}=\sigma(s_{ij})$, for each candidate edge $(i,j)\in E$,
where $s_{ij}$ is the decoder score for the candidate transition from $v_i$ to $v_j$;
$\sigma(\cdot)$ denotes the sigmoid activation function that maps the score to a probability; and 
$\hat{p}_{ij}$ is the predicted probability that the directed transition exists.}

\subsubsection{Auxiliary Widget-Attribute Reconstruction Objective}
As explained in Section~\ref{sec:motivation}, widget-trigger information should be modeled as an activity-independent signal.
However, the gravity-based decoder only predicts transition existence from activity embeddings $\mathbf{z}_i$ and $\mathbf{z}_j$, and does not directly use widget attributes as inputs.
To make widget-trigger information useful during model learning, an auxiliary objective for reconstructing widget features is used.

For each candidate transition edge $(i,j)\in E$ with an associated widget attribute vector $\mathbf{w}_{ij}$, 
a \textit{Multi-Layer Perceptron} (MLP)~\cite{deeplearningbook2016} is used  to predict a reconstructed attribute vector $\hat{\mathbf{w}}_{ij}$:
$\hat{\mathbf{w}}_{ij}=g_{\theta}\bigl([\mathbf{z}_i;\mathbf{z}_j]\bigr)$,
where $[\cdot;\cdot]$ denotes vector concatenation, and $g_{\theta}$ is trained jointly with the encoder and decoder.
This reconstruction is only applied to positive transition edges, to avoid learning the trigger attributes for non-transitions.
Formally, if $E^{\mathrm{attr}}=\{(i,j)\in E \mid y_{ij}=1\}$, then for each $(i,j)\in E^{\mathrm{attr}}$, the reconstruction compares $\hat{\mathbf{w}}_{ij}$ with the original $\mathbf{w}_{ij}$.

This encourages the learned activity representations to retain widget-trigger information.

\subsubsection{Learning for ATG Construction}
\label{sec:alpha_beta}
These three components constitute the backbone of \builder.
This section describes how \builder\ is trained, including:
(1) the training objectives and optimization procedure; and
(2) the decision rule for predicting transition existence.

\textit{(1) Training Objectives and Optimization.}
\builder\ aims to predict the existence of a candidate transition between two activities.
The primary task, therefore, remains transition prediction, with widget-attribute reconstruction serving as auxiliary regularization.
Two edge sets are used:
The \textit{prediction edge set} $E^{\mathrm{pred}}$ is used for supervised link prediction and evaluation; while 
the \textit{message-passing edge set} $E^{\mathrm{mp}}$ is only used by the GNN encoder for propagation.
This separation decouples representation learning from the supervised prediction space.
We use a smaller kNN graph ($k=20$) to define $E^{\mathrm{pred}}$, and a larger one ($k=60$) to define $E^{\mathrm{mp}}$:
If $E^{\mathrm{mp}}$ is unavailable, then $E^{\mathrm{pred}}$ is used.

Two objectives are used in \builder:
(1) For link prediction, a binary cross-entropy is applied on $E^{\mathrm{pred}}$:
$\mathcal{L}_{\mathrm{exist}}
=
-\frac{1}{|E^{\mathrm{pred}}|}
\sum_{(i,j)\in E^{\mathrm{pred}}}
\left[
y_{ij}\log \hat{p}_{ij}
+
(1-y_{ij})\log(1-\hat{p}_{ij})
\right]$.
(2) For widget-attribute reconstruction, MSE is applied on positive edges when edge attributes are enabled.
If edge attributes are enabled, then $\mathcal{L}_{\mathrm{attr}}$
is computed as
$\frac{1}{|E^{\mathrm{attr}}|}\sum_{(i,j)\in E^{\mathrm{attr}}}\left\|\hat{\mathbf{w}}_{ij}-\mathbf{w}_{ij}\right\|_2^2$;
if the edge feature is disabled, then $\mathcal{L}_{\mathrm{attr}}=0$.
The overall objective is
$\mathcal{L}=\alpha\mathcal{L}_{\mathrm{exist}}+\beta\mathcal{L}_{\mathrm{attr}}$,
where $\alpha$ and $\beta$ are two learnable parameters to balance the link-prediction loss and the widget-attribute reconstruction loss, respectively.

The ATGs obtained by \frontmatter\ can be noisy and incomplete~\cite{archidroid2023}.
To mitigate the noise in analysis-derived seed supervision, two noise-robust training mechanisms can be enabled during optimization:
(1) \emph{seed-consistency regularization}~\cite{seedconsistency2017}, which encourages prediction consistency under stochastic perturbations for the same candidate edge; and
(2) \emph{bootstrapping}~\cite{bootstrapping2015}, which mixes the seed labels with the model's current predictions to soften the influence of unreliable labels.

\textit{(2) Transition-Existence Decision Rule.}
After training, the model outputs a transition probability $\hat{p}_{ij}$ for each candidate edge $(i,j)\in E^{\mathrm{pred}}$.
Given a decision threshold $t$, \builder\ predicts the existence of a transition as $\hat{y}_{ij}=1$ iff $\hat{p}_{ij}\ge t$, and $\hat{y}_{ij}=0$, otherwise.
The default threshold of $t=0.5$ is a standard choice for binary classification~\cite{archidroid2023}.
However, in practice, a fixed threshold may not generalize well across tasks, datasets, and training conditions:
\builder\, therefore, selects a validation-based threshold $t^*$ through a grid search on the validation set during training (maximizing a validation criterion), and then evaluates this on the test set, aiming to avoid test-set leakage.

Based on these steps, the predicted ATG $\mathcal{G}_{\mathrm{pred}}$ is constructed by retaining those candidate edges for which $\hat{y}_{ij}=1$.

\section{Experimental Study} 
\label{sec:experimental-study}

This section presents the experimental design, including the research questions, experimental setup, and running environment.

\subsection{Research Questions} 
\label{sec:research-questions}

The following three research questions guided our experimental evaluation of \builder:

\begin{description}[leftmargin=0em, topsep=0pt, itemsep=0pt, parsep=0pt, partopsep=0pt]
    \item 
	    \textbf{RQ1 [Ablation Study]} \textit{What is the contribution of each core design choice to the \builder\ performance?}
	        (RQ1.1 for prediction-threshold selection;
	        RQ1.2 for multi-task loss weighting;
	        RQ1.3 for noise-robust training;
	        RQ1.4 for multi-dimensional feature fusion; and
	        \revp{RQ1.5 for sparse and balanced edge distributions}.)

    \item 
    \textbf{RQ2 [Effectiveness and Robustness Evaluation]} \textit{How effective and robust is \builder\ for real-world ATG construction?}
        (RQ2.1 for effectiveness comparison; and
        RQ2.2 for robustness under noisy seed supervision.)

    \item 
    \textbf{RQ3 [Usefulness Evaluation]} \textit{\revp{Can \builder-predicted ATGs improve the exploration effectiveness of automated GUI-exploration tools?}}
\end{description}

\subsection{Experimental Setup} 
\label{sec:experimental-setup}

\subsubsection{Setup for RQ1: Ablation Study}
\label{sec:setup-rq1}
This section describes the dataset preparation, the summary and embedding pipeline, and the ablation protocols for the experiments conducted to answer RQ1.

\begin{table*}
    \centering
    \scriptsize
    \setlength{\tabcolsep}{1.15mm}
    \caption{Dataset statistics for the \frontmatter\ ATGs.}
    \label{tab:frontmatter_allinone}
    \begin{tabular}{llrrrrrrrr}
        \hline
        \multirow{2}{*}{\textbf{Filter Stage}} &
        \multirow{2}{*}{\shortstack[l]{\textbf{Activity}\\[-2pt]\textbf{Naming Scene}}} &
        \multirow{2}{*}{\textbf{\#Apps}}&
        \multicolumn{3}{c}{\textbf{ATG Statistics}} &&
        \multicolumn{3}{c}{\textbf{Edge-set statistics}} \\
        \cline{4-6}\cline{8-10}
        & & & \textbf{\#Acts} & \textbf{\#Trans(w/)} & \textbf{\#Trans(w/o)}
        & & \textbf{\#Trans} & \textbf{\#Unlabeled} & \textbf{\#Edges} \\
        \hline
        
        \multirow{3}{*}{(1) All indexed apps}
        & Activity-suffixed
        & 12,983  & 106,483   & -- & -- & & 86,771     & -- & -- \\
        & Unconstrained 
        & 164,536 & 2,410,292 & -- & -- & & 14,459,375 & -- & -- \\
        \cline{2-10}
        & \textit{Total}
        & 177,519 & 2,516,775 & -- & -- & & 14,546,146 & -- & -- \\
        \hline
        
        \multirow{3}{*}{\shortstack[l]{(2) After size filtering\\[-2pt]($\#\mathrm{Acts}>5 \land \#\mathrm{Trans}>5$)}}
        & Activity-suffixed 
        & 3,000   & 42,865    & -- & -- & & 62,429     & -- & -- \\
        & Unconstrained 
        & 74,545  & 1,933,459 & -- & -- & & 14,232,343 & -- & -- \\
        \cline{2-10}
        & \textit{Total}
        & 77,545 & 1,976,324 & -- & -- & & 14,294,772 & -- & -- \\
        \hline
        
        \multirow{3}{*}{(3) Final sampled}
        & Activity-suffixed 
        & 2,500 & 36,478 & 26,005 & 22,847  & & 48,852  & 900,193  & 949,045 \\
        & Unconstrained 
        & 2,500 & 66,713 & 50,618 & 273,008 & & 323,626 & 2,660,402 & 2,984,028 \\
        \cline{2-10}
        & \textit{Total}
        & 5,000 & 103,191 & 76,623 & 295,855 & & 372,478 & 3,560,595 & 3,933,073 \\
        \hline
    \end{tabular}
    \parbox{\linewidth}{\scriptsize\emph{Note.}
    Counts are aggregated over apps.
    \#Trans(w/) and \#Trans(w/o) denote transitions with and without widget-trigger information, respectively.
    \#Unlabeled denotes non-transition edges, and $\mathrm{\#Edges}=\mathrm{\#Trans}+\mathrm{\#Unlabeled}$.}
\end{table*}

\textbf{Dataset Preparation:}
RQ1 used the \frontmatter\ dataset~\cite{frontmatter2021}, which provides static-analysis-derived seed transitions, and has been used in previous ATG prediction work~\cite{archidroid2023}.
All released ATGs were downloaded from the official dataset website~\cite{frontmatter_dataset}:
The same size-filtering principle as for \archidroid~\cite{archidroid2023} was then applied, and apps with more than five activities and more than five transitions were retained. 
Two activity-naming scenarios were considered: 
\textttsplie{*Activity}-suffixed, and unconstrained naming.
Filtering produced 77,545 candidate apps across the two scenarios:
Finally, 2500 apps were sampled per scenario (5000 total) for RQ1 (as shown in Table~\ref{tab:frontmatter_allinone}). 
Because the ATGs obtained by \frontmatter\ can be incomplete~\cite{frontmatter2021,archidroid2023}, they were used for ablation trends, rather than for real-world effectiveness claims.

\begin{table}
    \centering
    \caption{Embedding models used for textual features in \builder.}
    \label{TAB:embedding_model}
    \setlength\tabcolsep{1.8mm}
    \scriptsize
    \begin{tabular}{l l l c c c l}
        \hline
        \textbf{No.} & \textbf{Model name} & \textbf{Category} & \textbf{Hidden size} & \textbf{Vocab size} & \textbf{\#Layers} & \textbf{Used for} \\
        \hline
        1 & \texttt{bert-base-cased}~\cite{bert_base_cased}    & Word encoder    & 768 & 28,996 & 12 & Activity names and widget types \\
        2 & \texttt{all-MiniLM-L12-v2}~\cite{all_minilm_l12_v2}   & Sentence encoder & 384 & 30,522 & 12 & Activity and widget summaries \\
        \hline
    \end{tabular}
\end{table}

\textbf{Summary and Embedding Pipeline:}
GPT-4o~\cite{gpt4o} was used to generate one-sentence functionality summaries for both activities and widgets.
BERT~\cite{archidroid2023,tacdroid2025} was used to encode the raw activity identifiers and the widget types, and a sentence-embedding model~\cite{sentencebert2019} was used to encode the LLM-generated summaries.
Table~\ref{TAB:embedding_model} summarizes the embedding models and their roles in \builder.

\textbf{Ablation Protocols:}
To avoid a combinatorial explosion, a \textit{data protocol} and a \textit{model protocol} were defined for the RQ1 ablations:
\begin{itemize}[leftmargin=2em, topsep=0pt, itemsep=0pt, parsep=0pt, partopsep=0pt]
	\item 
	\textit{Data protocol.}
	The 5000 selected apps were split into train/validate/test subsets, with a \seqsplit{0.7/0.1/0.2} ratio.
	The experiments were repeated across three split seeds, with sampling-based settings repeated three times for each split.
	    
    \item 
    \textit{Model protocol.} 
    All variants were trained over 80 epochs, with a learning rate of $0.005$, weight decay of $1\mathrm{e}{-5}$, and batch size of 128. 
    The threshold $t^*$ was selected through a grid search over $[0.001, 0.999]$ on the validation set, to maximize the F1-score.
\end{itemize}

A controlled design was used, with one factor varied at a time, while the remaining factors remained fixed. 
RQ1.1 compared two prediction-threshold strategies: a fixed threshold ($0.5$) and a validation-based threshold ($t^*$) (Section~\ref{sec:alpha_beta}).
RQ1.2 studied multi-task loss weighting by comparing two strategies: 
Fixing $\alpha=1$ while learning $\beta$; and 
learning both $\alpha$ and $\beta$ (Section~\ref{sec:method}). 
To assess the contribution of the auxiliary widget-attribute reconstruction objective, this component was ablated by setting $\beta = 0$, and evaluating two variants: 
one with a fixed $\alpha = 1$ and another that treated $\alpha$ as a learnable parameter.
For RQ1.3, two optional noise-robust training mechanisms (Section~\ref{sec:alpha_beta}) were evaluated under four settings: 
base; 
seed-consistency only;
bootstrapping only; and 
the combination of seed-consistency and
bootstrapping.
RQ1.4 ablated activity features compared with widget features.
\revp{RQ1.5 considered two edge-distribution settings:
(1) \textit{no-sampling} retained all available non-transition candidate edges, and preserved the naturally sparse distribution in the real-world edge space; and
(2) \textit{balanced-sampling} sampled the same number of negative edges as positive transitions for each graph, resulting in a balanced distribution.}

\subsubsection{Setup for RQ2: Effectiveness and Robustness Evaluation}
This section introduces the benchmark preparation and protocols for evaluating the effectiveness and robustness of \builder.

\textbf{Benchmark Preparation:}
RQ2 involved evaluating \builder\ on a benchmark of 98 Android apps with manually checked ground-truth ATGs~\cite{atgemprical2025}.
\builder\ was compared with seven state-of-the-art (SOTA) baselines:
\monkey~\cite{monkey},
\stoat~\cite{stoat2017},
\ape~\cite{ape2019},
\qtest~\cite{qtesting2020},
\humanoid~\cite{humanoid2019},
\fastbot~\cite{fastbot22022}, and
\scenedroid~\cite{scenedroid2023}.
These tools were previously selected as benchmark ATG-construction tools~\cite{atgemprical2025}.

\textbf{Effectiveness-Evaluation Protocols:}
Two evaluation protocols were used to examine the effectiveness of \builder\ in constructing the ATGs:
\begin{itemize}[leftmargin=2em, topsep=0pt, itemsep=0pt, parsep=0pt, partopsep=0pt]
    \item 
    \textit{Non-finetuning (NF) protocol.} 
    The model trained on the RQ1 corpus (\frontmatter~\cite{frontmatter2021}) was directly evaluated on the 98-app benchmark set, without any adaptation.
    
    \item 
    \textit{Finetuning (FT) protocol.}
    The RQ1-pretrained model was finetuned on the training subset of the 98-app benchmark before evaluation.
\end{itemize}

\textbf{Robustness-Evaluation Protocols:}
RQ2.2 assessed the robustness of \builder\ by introducing false-positive noise in the seed supervision.
Dynamically-executed seed transitions were constructed using \ape~\cite{ape2019} and \stoat~\cite{stoat2017}:
These were considered the seed positives.
The $y^{\mathrm{seed}}_{ij}$ was $1$ iff $(i,j)$ appeared in these tool-discovered transitions, and $y^{\mathrm{seed}}_{ij}=0$, otherwise.
Accordingly, the seed-labeled candidate-edge sets were defined as
$E^{\mathrm{pos}}=\{(i,j)\in E \mid y^{\mathrm{seed}}_{ij}=1\}$ and
$E^{\mathrm{neg}}=\{(i,j)\in E \mid y^{\mathrm{seed}}_{ij}=0\}$.
During training, synthetic noise was injected into $\mathbf{y}^{\mathrm{seed}}$ to obtain $\mathbf{y}^{\mathrm{train}}$, with $\eta\in[0,1]$ denoting the noise rate (i.e., the proportion of seed labels that was corrupted).
There were two modes:
\begin{itemize}[leftmargin=2em, topsep=0pt, itemsep=0pt, parsep=0pt, partopsep=0pt]
    \item 
    \textit{FP-only noise.}
    If $k=\min\bigl(\operatorname{round}(\eta|E^{\mathrm{pos}}|),|E^{\mathrm{neg}}|\bigr)$, 
    then $E^{\mathrm{samp}}\subseteq E^{\mathrm{neg}}$ was uniformly sampled with $|E^{\mathrm{samp}}|=k$ and $y^{\text{train}}_{ij}=1$ iff $(i,j)\in E^{\mathrm{pos}}\cup E^{\mathrm{samp}}$, and $y^{\text{train}}_{ij}=0$, otherwise.

    \item 
    \textit{Uniform-flip noise.}
    The seed label was independently flipped for each edge $(i,j)$:
    $y^{\text{train}}_{ij}=1-y^{\text{seed}}_{ij}$ with probability $\eta$; and
    $y^{\text{train}}_{ij}=y^{\text{seed}}_{ij}$ with probability $1-\eta$.
\end{itemize}
This was done for $\eta \in \{0, 0.05, 0.10, \ldots, 1.00\}$ (with a step size of $0.05$).

\subsubsection{Setup for RQ3: Usefulness Evaluation}
\revp{To evaluate the practical usefulness of \builder-predicted ATGs, we examined whether they could improve existing automated GUI-exploration tools.}
\revp{Three representative tools were considered:
\monkey~\cite{monkey}, \ape~\cite{ape2019}, and \fastbot~\cite{fastbot22022}.
For each tool, both its original version and its \builder-guided variant were evaluated:}
\revp{The \textit{original tools} used their native exploration strategies and official/default configurations~\cite{monkey,ape2019,fastbot22022}.}
\revp{Following the ATG-guided approach by \citet{archidroid2023}, the \textit{\builder-guided variants} used the \builder-predicted ATG as an external episode-level testing scheduler~\cite{archidroid2023}, without modifying the internal exploration logic of the original tools.
Specifically:
(1) Each activity degree was calculated as the sum of the incoming and outgoing transitions of the activity node.
    The testing budget was then allocated proportionally to these activities and split into multiple episodes.
(2) Before each episode, the testing scheduler selected an activity with remaining testing budget, preferably one that appeared in the predicted ATG but had not yet been observed during exploration.
(3) If possible, the scheduler then directly launched the selected activity:
    Otherwise, it fell back to the nearest launchable preceding activity in the predicted ATG, or the app entry activity.
(4) The execution ran unchanged until the end of the episode budget, after which the next episode was initiated.
    \builder\ was used only as a lightweight navigation model for activity selection and time allocation;
    the UI-event generation remained fully controlled by the original tools.}

Five apps were randomly selected from the test set for RQ2.1 as the \textit{Apps Under Test} (AUTs).
\revp{Each strategy was evaluated on the same AUTs under identical device settings, with ten one-hour runs:
Before each run, the app state was reset to eliminate interference from previous executions.}

\subsection{Evaluation Metrics}
\label{sec:evaluation-metrics}

This section presents the evaluation metrics used in our experiments.

\subsubsection{Metrics for RQ1 and RQ2}
\label{sec:rq1-rq2-metrics}
For each app, $E$ denotes the candidate edge set, and $y_{ij}\in\{0,1\}$ is the reference label for each $(i,j)\in E$.
Our model outputs a transition probability $\hat{p}_{ij}\in[0,1]$, and predicts $\hat{y}_{ij}=1$ iff $\hat{p}_{ij}\ge t$, and $\hat{y}_{ij}=0$, otherwise. 
If $Y=\{(i,j)\in E \mid y_{ij}=1\}$ and $\hat{Y}=\{(i,j)\in E \mid \hat{y}_{ij}=1\}$, then:
$\mathrm{Precision}=\frac{|Y\cap \hat{Y}|}{|\hat{Y}|}$, 
$\mathrm{Recall}=\frac{|Y\cap \hat{Y}|}{|Y|}$, and 
$\text{F1-score}=\frac{2\times \mathrm{Precision}\times \mathrm{Recall}}{\mathrm{Precision}+\mathrm{Recall}}$~\cite{tacdroid2025,archidroid2023,harmonyos2024}.
The same metrics were used for RQ1 and RQ2, but with different references: 
RQ1 used \frontmatter's static-analysis-derived labels~\cite{frontmatter2021}, whereas RQ2 used manually-checked ground-truth ATGs~\cite{atgemprical2025}.
The output from a GUI-testing baseline tool is a discrete transition set $\hat{Y}_{\text{tool}}$.
\revp{All ATGs were aligned at the activity-transition level before computing the metrics. 
Nodes were normalized by activity names, and edges were compared as source-destination activity pairs.}

\subsubsection{Metrics for RQ3}
\label{sec:rq3-metrics}
Given the ground-truth ATG with activity set $A$ and transition set $T$, at time $\tau$, the visited activities are denoted as $A_\tau\subseteq A$, and the observed ground-truth transitions are denoted as $T_\tau\subseteq T$.
Two coverage metrics~\cite{archidroid2023,harmonyos2024} were used:
$\mathrm{AC}(\tau)=\frac{|A_\tau|}{|A|}$ (activity coverage) and $\mathrm{TC}(\tau)=\frac{|T_\tau|}{|T|}$ (transition coverage).
Two process metrics~\cite{fillintheblank2023,harmonyos2024} were also recorded: 
$\mathrm{UN}(\tau)=|U_\tau|$ (the number of distinct UI states observed so far) and $\mathrm{EN}(\tau)$ (the total number of injected UI events up to time $\tau$).
Progress efficiency was measured with per-event change rates~\cite{guitar2014,toller2021}:
$\mathrm{ACR}(\tau)=\frac{1}{\mathrm{EN}(\tau)}\sum_{i=1}^{\mathrm{EN}(\tau)}\mathbb{I}\big[a_i\neq a_{i-1}\big]$, and
$\mathrm{UCR}(\tau)=\frac{1}{\mathrm{EN}(\tau)}\sum_{i=1}^{\mathrm{EN}(\tau)}\mathbb{I}\big[u_i\neq u_{i-1}\big]$,
where $a_i$ and $u_i$ are the activity and UI state observed at step $i$, and $\mathbb{I}[\cdot]$ is the indicator function.

\subsection{Running Environment}
\label{sec:running-environment}

All model-related experiments were conducted on a Linux server with an NVIDIA RTX 4090 GPU (24\,GB), 16 vCPUs (Intel Xeon Platinum 8358P @ 2.60\,GHz), and 120\,GB RAM, using Python 3.10 (Ubuntu 22.04), PyTorch 2.1.2, and CUDA 11.8. 
\revp{Online app runs were executed on Android emulators running Android 11 (API 30), each configured with four virtual CPU cores and 2\,GB RAM.}

\section{Experimental Results} 
\label{sec:experimental-results}

This section reports the results for each research question in Section~\ref{sec:research-questions}.
All results are summarized as mean $\pm$ standard deviation over three split seeds under a fixed threshold ($@0.5$), and a validation-selected threshold ($@t^*$).
Wavy underlines indicate the best value within each row group, and shaded cells indicate the overall best value in the table.

\subsection{Answer to RQ1: Ablation Study} 
\label{sec:res-rq1}

This section reports the ablation results for RQ1, quantifying the contribution of each core design choice in \builder.

\begin{table*}
    \centering
    \scriptsize
    \setlength{\tabcolsep}{1.78mm}
    \caption{RQ1.1: Effect of prediction-threshold selection.}
    \label{tab:rq11-threshold}
    \begin{tabular}{@{}c c c c c c c c c c c@{}}

        \hline
        \textbf{ID} &
        \textbf{Encoder} &
        &
        \textbf{P (@0.5)} &
        \textbf{R (@0.5)} &
        \textbf{F1 (@0.5)} &
        &
        \textbf{P (@$t^*$)} &
        \textbf{R (@$t^*$)} &
        \textbf{F1 (@$t^*$)} &
        \textbf{$t^*$ Value} \\
        \hline

        \textbf{E0} & GIN &
        & \hlo{\rwavo{$0.9057$}}$_{\pm 0.0121}$ 
        & \hlo{\rwavo{$0.9139$}}$_{\pm 0.0508}$ 
        & \hlo{\rwavo{$0.9072$}}$_{\pm 0.0321}$
        &
        & \hlb{\rwavb{$0.8995$}}$_{\pm 0.0131}$ 
        & \hlb{\rwavb{$0.9330$}}$_{\pm 0.0230}$ 
        & \hlb{\rwavb{$0.9158$}}$_{\pm 0.0177}$ 
        & $0.4535_{\pm 0.0423}$ \\
        \hline
    \end{tabular}
\end{table*}

\subsubsection{Answer to RQ1.1: Prediction-Threshold Comparison}
Table~\ref{tab:rq11-threshold} presents the results for two decision thresholds.
The results indicate that \textbf{selecting the decision threshold on the validation set yields a better performance than using a fixed threshold.}
Compared with $@0.5$, using $@t^*$ increased recall (0.9139 to 0.9330) with only a slight precision decrease (0.9057 to 0.8995), resulting in a higher F1-score (0.9072 to 0.9158).
The selected $t^*$ was around 0.45, indicating that a fixed threshold (e.g., 0.5) may not be optimal for ATG construction.
This is as expected, because the prediction-probability distribution can vary with the ATG structure and the quality of the seed supervision.
A validation-selected $t^*$ provides a practical decision rule for ATG construction in this specific prediction task.

\rqsummary{1.1}
{We recommend that the prediction-threshold selection be treated as part of the model configuration, and that it be tuned for the encoder (using a validation-selected $t^*$), rather than using a fixed value.}

\begin{table*}
    \centering
    \scriptsize
    \setlength{\tabcolsep}{0.87mm}
    \caption{RQ1.2: Effect of dynamic loss-weight learning strategies.}
    \label{tab:rq12-loss-balance}   
    \begin{tabular}{@{}c c c c c c c c c c c@{}}

        \hline
        \textbf{ID} &
        \textbf{Learned $\bm{\alpha/\beta}$ Value} &
        &
        \textbf{P (@0.5)} &
        \textbf{R (@0.5)} &
        \textbf{F1 (@0.5)} &
        &
        \textbf{P (@$t^*$)} &
        \textbf{R (@$t^*$)} &
        \textbf{F1 (@$t^*$)} &
        \textbf{$t^*$ Value} \\
        \hline

        \textbf{L1} & $0.6667_{\pm 0.0568}/0.4915_{\pm 0.0022}$ &
        & $0.8975_{\pm 0.0218}$ 
        & \hlo{\rwavo{$0.9305$}}$_{\pm 0.0284}$ 
        & \hlo{\rwavo{$0.9135$}}$_{\pm 0.0197}$
        &
        & $0.8942_{\pm 0.0163}$ 
        & $0.9349_{\pm 0.0306}$ 
        & $0.9139_{\pm 0.0205}$ 
        & $0.4720_{\pm 0.0596}$ \\
        
        \textbf{L2} & $1.0000_{\pm 0.0000}/0.4909_{\pm 0.0012}$ &
        & \hlo{\rwavo{$0.9057$}}$_{\pm 0.0120}$ 
        & $0.9138_{\pm 0.0508}$ 
        & $0.9072_{\pm 0.0320}$
        &
        & \hlb{\rwavb{$0.8995$}}$_{\pm 0.0130}$ 
        & $0.9330_{\pm 0.0230}$ 
        & \hlb{\rwavb{$0.9158$}}$_{\pm 0.0177}$ 
        & $0.4535_{\pm 0.0422}$ \\
        
        \textbf{L3} & $0.7572_{\pm 0.0623}/0.0000_{\pm 0.0000}$ &
        & $0.8953_{\pm 0.0155}$ 
        & $0.9264_{\pm 0.0224}$ 
        & $0.9104_{\pm 0.0143}$
        &
        & $0.8938_{\pm 0.0173}$ 
        & $0.9302_{\pm 0.0156}$ 
        & $0.9115_{\pm 0.0129}$ 
        & $0.4849_{\pm 0.0812}$ \\
        
        \textbf{L4} & $1.0000_{\pm 0.0000}/0.0000_{\pm 0.0000}$ &
        & $0.8893_{\pm 0.0204}$ 
        & $0.9296_{\pm 0.0178}$ 
        & $0.9089_{\pm 0.0150}$
        &
        & $0.8820_{\pm 0.0199}$ 
        & \hlb{\rwavb{$0.9397$}}$_{\pm 0.0145}$ 
        & $0.9098_{\pm 0.0146}$ 
        & $0.4429_{\pm 0.0901}$ \\
        
        \hline
    \end{tabular}
\end{table*}

\subsubsection{Answer to RQ1.2: Multi-task Loss Weighting}
Table~\ref{tab:rq12-loss-balance} presents the results for four loss-weight configurations:
(L1) learning both $\alpha$ and $\beta$;
(L2) fixing $\alpha=1$ while learning $\beta$; and
two diagnostic variants that disable the widget-attribute reconstruction objective by setting $\beta=0$,
with either a learnable $\alpha$ (L3) or a fixed $\alpha=1$ (L4).
Based on these results, we have the following observations:
\begin{itemize}[leftmargin=2em, topsep=0pt, itemsep=0pt, parsep=0pt, partopsep=0pt]    
    \item 
    \textbf{Enabling the widget-attribute reconstruction objective yielded better performance.}
    The best results were obtained with reconstruction enabled (L2 under $@t^*$ and L1 under $@0.5$).
    Disabling reconstruction ($\beta=0$; L3/L4) led to a small decrease in the F1-score under $@0.5$, suggesting that widget-attribute reconstruction provided an auxiliary regularization benefit rather than a dominant learning signal: 
    This is consistent with our original intuition.

    \item 
    \textbf{Treating $\beta$ as learnable was more robust than fixing it.}
    When the widget-trigger information was identified, the reconstruction objective could strengthen the \builder\ performance.
    However, when the information was weak, such as with a placeholder embedding, overly emphasizing the reconstruction could risk over-regularization.
    Learning $\beta$ allowed \builder\ to
    automatically reduce the weights ($\mathcal{L}_{\mathrm{attr}}$) when the reconstructed attributes were less reliable, and increase them when the attributes were informative:
    This increased the training stability.
\end{itemize}

\rqsummary{1.2}{Multi-task loss weighting had a practical impact: it mainly changed the distribution of predicted probabilities rather than changing the overall ability of \builder\ to distinguish transitions from non-transitions.
We recommend enabling the widget-attribute reconstruction objective as a default setting.}

\begin{table*}
    \centering
    \scriptsize
    \setlength{\tabcolsep}{1.05mm}
    \caption{RQ1.3: Effect of seed-consistency and bootstrapping on noise robustness.}
    \label{tab:rq13-noise-robust}
    \begin{tabular}{@{}c c c c c c c c c c c@{}}

        \hline
        \textbf{ID} &
        \textbf{Noise-Robust Setting} &
        &
        \textbf{P (@0.5)} &
        \textbf{R (@0.5)} &
        \textbf{F1 (@0.5)} &
        &
        \textbf{P (@$t^*$)} &
        \textbf{R (@$t^*$)} &
        \textbf{F1 (@$t^*$)} &
        \textbf{$t^*$ Value} \\
        \hline
        
        \textbf{N1} & Base &
        & \hlo{\rwavo{$0.9057$}}$_{\pm 0.0120}$ 
        & $0.9138_{\pm 0.0508}$ 
        & $0.9072_{\pm 0.0320}$
        &
        & \hlb{\rwavb{$0.8995$}}$_{\pm 0.0130}$ 
        & $0.9330_{\pm 0.0230}$ 
        & $0.9158_{\pm 0.0177}$ 
        & $0.4535_{\pm 0.0422}$ \\
        
        \textbf{N2} & Seed-consis. only &
        & $0.8999_{\pm 0.0118}$ 
        & \hlo{\rwavo{$0.9430$}}$_{\pm 0.0162}$ 
        & \hlo{\rwavo{$0.9189$}}$_{\pm 0.0144}$
        &
        & $0.8985_{\pm 0.0128}$ 
        & \hlb{\rwavb{$0.9461$}}$_{\pm 0.0263}$ 
        & \hlb{\rwavb{$0.9194$}}$_{\pm 0.0196}$ 
        & $0.4602_{\pm 0.0224}$ \\
        
        \textbf{N3} & Boots. only &
        & $0.9025_{\pm 0.0129}$ 
        & $0.9246_{\pm 0.0127}$ 
        & $0.9142_{\pm 0.0120}$
        &
        & $0.8989_{\pm 0.0142}$ 
        & $0.9399_{\pm 0.0098}$ 
        & $0.9190_{\pm 0.0066}$ 
        & $0.4343_{\pm 0.0487}$ \\
        
        \textbf{N4} & Seed-consis. + boots. &
        & $0.8971_{\pm 0.0177}$ 
        & $0.9231_{\pm 0.0194}$ 
        & $0.9102_{\pm 0.0182}$
        &
        & $0.8955_{\pm 0.0173}$ 
        & $0.9273_{\pm 0.0315}$ 
        & $0.9106_{\pm 0.0224}$ 
        & $0.4417_{\pm 0.0388}$ \\
        
        \hline
    \end{tabular}
\end{table*}

\subsubsection{Answer to RQ1.3: Noise-Robustness Training}
Table~\ref{tab:rq13-noise-robust} presents the results for using seed-consistency and bootstrapping mechanisms under unreliably-labeled static-seed supervision.
Based on the results, we have the following observations:
\begin{itemize}[leftmargin=2em, topsep=0pt, itemsep=0pt, parsep=0pt, partopsep=0pt]
    \item 
    \textbf{Seed-consistency achieved the most consistent improvements.}
    Compared with the base setting (N1), seed-consistency (N2) had an increased F1-score of from 0.9072 to 0.9189 under $@0.5$, and from 0.9158 to 0.9194 under $@t^*$:
    This suggests that seed-consistency encourages the model to produce consistent predictions, even under perturbations, thereby reducing sensitivity to false positives/negatives labels in static-seed supervision.

    \item 
    \textbf{Bootstrapping also improved performance, but less than with seed-consistency.}
    Bootstrapping uses a mixture of labels and model predictions, and improved the F1-scores to 0.9142 ($@0.5$) and 0.9190 ($@t^*$):
    This suggests that once the model has learned a reasonably good decision boundary, then its predictions could soften the impact of incorrect seed labels.

    \item 
    \textbf{Combining the two mechanisms did not further improve the performance.}
    Combining seed consistency and bootstrapping (N4) degraded the F1-scores to around 0.910, under both thresholds.
    A possible explanation for this is over-regularization:
    Seed-consistency can make the model conservative and reduce positive predictions, which increases false negatives and lowers recall. 
    Bootstrapping can amplify early prediction errors by feeding them back into training, which increases false positives and lowers precision.
\end{itemize}

\rqsummary{1.3}{Because static-analysis-derived seeds may contain false positives/negatives~\cite{archidroid2023,improvestatic2026}, we recommend enabling seed-consistency as the default robustness mechanism, and using bootstrapping cautiously, especially when combining the two.}

\begin{table*}
    \centering
    \scriptsize
    \setlength{\tabcolsep}{0.9mm}
    %\renewcommand{\arraystretch}{0.73}
    %\captionsetup{skip=2pt}
    \caption{RQ1.4: Effect of multi-dimensional feature fusion.}
    \label{tab:rq14-feature-fusion}
    \begin{tabular}{@{}c c c c c c c c c c c@{}}
        % \hline
        % \multirowcell{2}{\textbf{ID}} &
        % \multirowcell{2}{\textbf{Variant}} &
        % &
        % \multicolumn{3}{c}{$\bm{@0.5}$} &
        % \multicolumn{1}{c}{} &
        % \multicolumn{4}{c}{$\mathit{\bm{@t^*}}$} \\
        % \cline{4-6}\cline{8-11}
        % & & &
        % \textbf{Precision} & \textbf{Recall} & \textbf{F1-score} &
        % &
        % \textbf{Precision} & \textbf{Recall} & \textbf{F1-score} & \textbf{$t^*$ Value} \\
        % \hline

        \hline
        \textbf{ID} &
        \textbf{Variant} &
        &
        \textbf{P (@0.5)} &
        \textbf{R (@0.5)} &
        \textbf{F1 (@0.5)} &
        &
        \textbf{P (@$t^*$)} &
        \textbf{R (@$t^*$)} &
        \textbf{F1 (@$t^*$)} &
        \textbf{$t^*$ Value} \\
        \hline
        
        \multicolumn{11}{l}{\textit{\textbf{Baseline and Complete Setting}}} \\
        \hline
        \textbf{T1} & Baseline &
        & $0.8879_{\pm 0.0160}$ & $0.9273_{\pm 0.0345}$ & $0.9028_{\pm 0.0101}$
        & 
        & $0.8814_{\pm 0.0145}$ & $0.9232_{\pm 0.0179}$ & $0.9051_{\pm 0.0202}$ & $0.4482_{\pm 0.0290}$ \\
        \textbf{T2} & Complete &
        & \rwavo{$0.8947$}$_{\pm 0.0162}$ 
        & \rwavo{$0.9305$}$_{\pm 0.0284}$ 
        & \rwavo{$0.9135$}$_{\pm 0.0197}$
        & 
        & \rwavb{$0.8942$}$_{\pm 0.0163}$ 
        & \rwavb{$0.9349$}$_{\pm 0.0306}$ 
        & \rwavb{$0.9139$}$_{\pm 0.0205}$ 
        & $0.4720_{\pm 0.0596}$ \\
        \hline
        
        \multicolumn{11}{l}{\textit{\textbf{Widget-Feature Ablations (keeping Activity Name and Summary)}}} \\\hline
        \textbf{T3} & No Widget Features &
        & $0.8951_{\pm 0.0167}$ & $0.9288_{\pm 0.0281}$ & $0.9108_{\pm 0.0258}$
        & 
        & \rwavb{$0.8901$}$_{\pm 0.0240}$ & $0.9371_{\pm 0.0226}$ & $0.9114_{\pm 0.0268}$ & $0.4691_{\pm 0.0170}$ \\
        \textbf{T4} & Widget: Name only &
        & \rwavo{$0.9040$}$_{\pm 0.0130}$ & $0.9241_{\pm 0.0362}$ & \rwavo{$0.9139$}$_{\pm 0.0126}$
        & 
        & $0.8889_{\pm 0.0313}$ & \rwavb{$0.9448$}$_{\pm 0.0129}$ & \rwavb{$0.9154$}$_{\pm 0.0202}$ & $0.4322_{\pm 0.1015}$ \\
        \textbf{T5} & Widget: Summary only &
        & $0.8921_{\pm 0.0148}$ & \rwavo{$0.9304$}$_{\pm 0.0206}$ & $0.9104_{\pm 0.0265}$
        & 
        & $0.8867_{\pm 0.0216}$ & $0.9348_{\pm 0.0156}$ & $0.9095_{\pm 0.0253}$ & $0.4648_{\pm 0.1166}$ \\
        \hline
        
        \multicolumn{11}{l}{\textit{\textbf{Activity-Feature Ablations (keeping Widget Name and Summary)}}} \\\hline
        \textbf{T6} & Activity: Name only &
        & $0.8783_{\pm 0.0127}$ & \hlo{\rwavo{$0.9677$}}$_{\pm 0.0449}$ & $0.9171_{\pm 0.0362}$
        & 
        & $0.8510_{\pm 0.1262}$ & \hlb{\rwavb{$0.9682$}}$_{\pm 0.0440}$ & $0.9021_{\pm 0.0797}$ & $0.5454_{\pm 0.0980}$ \\
        \textbf{T7} & Activity: Summary only &
        & \hlo{\rwavo{$0.9183$}}$_{\pm 0.0092}$ 
        & $0.9412_{\pm 0.0166}$ 
        & \hlo{\rwavo{$0.9295$}}$_{\pm 0.0128}$
        & 
        & \hlb{\rwavb{$0.9108$}}$_{\pm 0.0109}$ 
        & $0.9563_{\pm 0.0064}$ 
        & \hlb{\rwavb{$0.9328$}}$_{\pm 0.0048}$ 
        & $0.3952_{\pm 0.0562}$ \\
        \textbf{T8} & Activity: Simple name &
        & $0.9088_{\pm 0.0145}$ & $0.9175_{\pm 0.0195}$ & $0.9130_{\pm 0.0171}$
        & 
        & $0.9001_{\pm 0.0138}$ & $0.9349_{\pm 0.0194}$ & $0.9160_{\pm 0.0188}$ & $0.4495_{\pm 0.0302}$ \\
        \textbf{T9} & Activity: Name w/o suffix &
        & $0.8879_{\pm 0.0172}$ & $0.9411_{\pm 0.0270}$ & $0.9136_{\pm 0.0119}$
        & 
        & $0.8729_{\pm 0.0202}$ & $0.9522_{\pm 0.0133}$ & $0.9105_{\pm 0.0198}$ & $0.4154_{\pm 0.1213}$ \\
        \hline
    \end{tabular}
    
\end{table*}

\subsubsection{Answer to RQ1.4: Multi-dimensional Feature Fusion}
Table~\ref{tab:rq14-feature-fusion} presents the results related to how different feature groups contribute to transition prediction. 
Based on these results, we have the following observations:
\begin{itemize}[leftmargin=2em, topsep=0pt, itemsep=0pt, parsep=0pt, partopsep=0pt]
    \item 
    \textbf{The fused features improved overall performance more than the identifier-only feature.}
    Overall, the complete setting improved over the baseline (T2 vs. T1): 
    Under $@0.5$, the F1-score rose from 0.9028 to 0.9135; while under $@t^*$, the performance remained comparable (0.9051 vs. 0.9139), indicating that richer features improved the performance. 
    This suggests that adding semantic features beyond raw activity identifiers can achieve a better representation for ATG construction.

    \item 
    \textbf{Activity features had the strongest impact.}
    Among all variants, using activity summaries alone (T7) achieved the best results (0.9295 under $@0.5$ and 0.9328 under $@t^*$). 
    This indicates that the most reliable information across apps was the semantic functionality of the activities:
    Raw activity identifiers were developer-chosen and often not standardized. 
    In contrast, the LLM-generated activity summaries provided a normalized semantic description that compressed layout and textual information into a functional representation.

    \item 
    \textbf{Widget-trigger information provided auxiliary improvements.}
    The widget-feature ablations (T3 to T5) yielded only slight improvements, suggesting that widget features mainly provide auxiliary support, with the primary gain coming from the activity semantics.
    Nevertheless, the widget-trigger information can still be useful, as it preserves the triggering context:
    This can help to distinguish multiple widget-triggered routes that previous work treated as a single activity-pair edge~\cite{archidroid2023}.
    However, the widget-trigger information may often be incomplete in the static metadata:
    Static extraction may miss runtime-dependent transitions that are realized through indirect control flows
    ---
    such as menus, fragments, and navigation components
    ---
    that did not require any explicit widget-based event to trigger them. 
    This limits the standalone benefit of widget-trigger information and makes it more suitable as an auxiliary training signal.
\end{itemize}

\rqsummary{1.4}{Feature fusion improved transition prediction, with the largest gain coming from activity functionality representations (LLM-generated summaries). 
We recommend using widget-trigger information as an auxiliary training signal.}

\subsubsection{\revp{Answer to RQ1.5: Sparse and Balanced Edge Distributions}}

\begin{table*}
    \color{black}
    \centering
    \scriptsize
    \setlength{\tabcolsep}{1.25mm}
    \renewcommand{\arraystretch}{0.85}
    \caption{\revp{RQ1.5: Effect of edge-distribution settings.}}
    \label{tab:rq15-sampling}
    \begin{tabular}{@{}c c c c c c c c c c c@{}}

        \hline
        \textbf{ID} &
        \textbf{Edge Distribution} &
        &
        \textbf{P (@0.5)} &
        \textbf{R (@0.5)} &
        \textbf{F1 (@0.5)} &
        &
        \textbf{P (@$t^*$)} &
        \textbf{R (@$t^*$)} &
        \textbf{F1 (@$t^*$)} &
        \textbf{$t^*$ Value} \\
        \hline

        \textbf{S1} & No Sampling &
        & $0.7614_{\pm 0.1406}$
        & $0.8234_{\pm 0.0380}$
        & $0.7883_{\pm 0.0964}$
        &
        & $0.7628_{\pm 0.1377}$
        & $0.8194_{\pm 0.0452}$
        & $0.7876_{\pm 0.0973}$
        & $0.5241_{\pm 0.0512}$ \\

        \textbf{S2} & Balanced Sampling &
        & \hlo{\rwavo{$0.9199$}}$_{\pm 0.0181}$
        & \hlo{\rwavo{$0.9425$}}$_{\pm 0.0202}$
        & \hlo{\rwavo{$0.9310$}}$_{\pm 0.0191}$
        &
        & \hlb{\rwavb{$0.9115$}}$_{\pm 0.0220}$
        & \hlb{\rwavb{$0.9543$}}$_{\pm 0.0143}$
        & \hlb{\rwavb{$0.9323$}}$_{\pm 0.0184}$
        & $0.4166_{\pm 0.0253}$ \\
        \hline
    \end{tabular}
\end{table*}

\revp{Table~\ref{tab:rq15-sampling} shows the results for two edge-distribution settings for evaluation, based on which, we have the following observations:}
\begin{itemize}[leftmargin=2em, topsep=0pt, itemsep=0pt, parsep=0pt, partopsep=0pt]
    \item
    \revp{\textbf{The no-sampling setting was substantially more insufficient.}
    Without negative sampling, the F1-score dropped to 0.7883 under $@0.5$, and 0.7876 under $@t^*$.
    This was expected, because the candidate edge space is highly imbalanced, with far more non-transition pairs than true activity transitions.
    In such sparse edge spaces, training with all negative candidates can make the model biased toward the majority (non-transition) class, weakening the learning signal from positive transitions.
    The no-sampling setting reflects the difficulty introduced by realistic edge imbalance, but it also makes model training more sensitive to the overwhelming number of negative edges.}

    \item
    \revp{\textbf{Balanced sampling produced more stable transition prediction.}
    With a balance between positive and negative samples, \builder\ achieved F1-scores of 0.9310 under $@0.5$, and 0.9323 under $@t^*$.
    The balanced setting thus provided a controlled evaluation protocol for learning to distinguish transitions from non-transitions.}
\end{itemize}

\rqsummary{1.5}{\revp{Sparse no-sampling setting was substantially insufficient, due to non-transition pairs dominating the candidate edge space.
Balanced sampling was more stable and controlled for ATG-link prediction, and is therefore used as the default setting for the controlled experiments.}}

\subsection{Answer to RQ2: Effectiveness and Robustness Evaluation}
\label{sec:res-rq2}

This section reports on \builder's effectiveness and robustness evaluation results on real-world apps with ground-truth ATGs.

\subsubsection{Answer to RQ2.1: Effectiveness Comparison}
\label{sec:answer2.1}

\revp{\builder\ was evaluated under two model-use settings on the 98-app benchmark set:
\begin{itemize}[leftmargin=2em, topsep=0pt, itemsep=0pt, parsep=0pt, partopsep=0pt]
    \item 
    \textit{Non-finetuning (\builder-NF)} reused the best model trained on the 5000-app corpus (Section~\ref{sec:res-rq1}).
    To estimate the offline training cost, the model was trained under three data-split seeds and three training seeds, resulting in nine runs.
    These runs took 38.97 minutes in total, with an average of $3.73~(\pm 0.03)$ minutes per run.

    \item
    \textit{Fine-tuning (\builder-FT)} further fine-tuned the pretrained model on the 98-app benchmark set.
    This step introduced little overhead, taking only 18 seconds across the nine runs, with an average of $1.6~(\pm 0.50)$ seconds per run.
\end{itemize}}

\begin{table*}
    \caption{RQ2.1: Generalization performance on the 98-app test set.}
    \label{tab:rq2_generalization_baselines}
    \centering
    \scriptsize
    \setlength{\tabcolsep}{8mm}
    \begin{tabular}{@{}lcccl@{}}
        \hline
        \textbf{Method} & \textbf{Precision} & \textbf{Recall} & \textbf{F1-score} &\textbf{Threshold} \\
        \hline
        \builder-NF @ 0.5      & \rwavg{$0.7567$}$_{\pm 0.0385}$ & $0.8169_{\pm 0.0482}$ & $0.7840_{\pm 0.0015}$ & $0.5$\\
        \builder-NF @ $t^*$    & $0.7214_{\pm 0.0245}$ & $0.9242_{\pm 0.0631}$ & $0.8094_{\pm 0.0249}$ & $0.2614_{\pm 0.1480}$\\
        \builder-FT @ 0.5      & $0.7449_{\pm 0.0429}$ & $0.8496_{\pm 0.0527}$ & $0.7918_{\pm 0.0042}$ & $0.5$\\
        \builder-FT @ $t^*$    & $0.7148_{\pm 0.0161}$ & \hlg{\rwavg{$0.9363$}}$_{\pm 0.0506}$ & \hlg{\rwavg{$0.8101$}}$_{\pm 0.0240}$ & $0.3034_{\pm 0.1785}$\\\hline
        
        \humanoid   & $0.6716_{\pm 0.0274}$ & $0.3109_{\pm 0.0868}$ & $0.4206_{\pm 0.0883}$ & -- \\
        \monkey     & $0.4559_{\pm 0.0806}$ & $0.1899_{\pm 0.0269}$ & $0.2669_{\pm 0.0329}$ & --\\
        \qtest      & \hlg{\rwavg{$0.9263$}}$_{\pm 0.0244}$ & $0.4429_{\pm 0.1231}$ & $0.5929_{\pm 0.1101}$ & --\\
        \stoat      & $0.8734_{\pm 0.0480}$ & $0.4650_{\pm 0.0507}$ & $0.6066_{\pm 0.0548}$ & --\\
        \ape        & $0.8481_{\pm 0.0378}$ & \rwavg{$0.5365$}$_{\pm 0.0393}$ & \rwavg{$0.6560$}$_{\pm 0.0235}$ & --\\
        \fastbot    & $0.8023_{\pm 0.0168}$ & $0.5279_{\pm 0.2202}$ & $0.6199_{\pm 0.1477}$ & --\\
        \scenedroid & $0.4222_{\pm 0.1347}$ & $0.0193_{\pm 0.0063}$ & $0.0367_{\pm 0.0115}$ & --\\
        \hline
    \end{tabular}
\end{table*}

\revp{These two settings were then compared against the selected baselines.}
Table~\ref{tab:rq2_generalization_baselines} presents the effectiveness-evaluation results for \builder\ on the 98-app benchmark.
The table includes a comparison with seven baselines that output discrete ATGs.
Based on these results, we have the following observations:
\begin{itemize}[leftmargin=2em, topsep=0pt, itemsep=0pt, parsep=0pt, partopsep=0pt]
    \item 
    \textbf{\builder\ achieved the best overall F1-score by substantially improving recall while keeping precision competitive.}
    The best configuration was \builder-FT@$t^*$, which achieved the highest recall (0.9363) and the highest F1-score (0.8101).
    Compared with tool baselines, \builder\ consistently recovered more ground-truth transitions while maintaining reasonable precision.

    \item 
    \textbf{Finetuning improved the effectiveness consistently, suggesting a domain gap between pretraining and the target benchmark.}
    For both thresholds, \builder-FT outperformed \builder-NF in both recall and F1-score:
    This indicates that the pretrained model did not perfectly match the 98-app benchmark, but that finetuning helped to adapt it to the target-domain transition patterns.

    \item 
    \textbf{Validation-based thresholding increased recall at the cost of precision, and worked particularly well after finetuning.}
    Switching from $@0.5$ to $@t^*$ increased the recall for both \builder-NF (0.8169 to 0.9242) and \builder-FT (0.8496 to 0.9363).
    This suggests that thresholding mainly controls the precision-recall trade-off, and that fine-tuning produces more stable probability scores on the target benchmarks, making $t^*$ more effective.
    
    \item 
    \textbf{Tool baselines were generally high-precision but low-recall under limited exploration budgets.}
    Among the tools, \qtest\ achieved the highest precision (0.9263) but had a much lower recall (0.4429), resulting in an F1-score well below that of \builder.
    Other tools showed a similar pattern, reflecting the inherent incompleteness of runtime exploration under limited budgets.
\end{itemize}

\begin{figure}
    \centering
    \includegraphics[width=\linewidth]{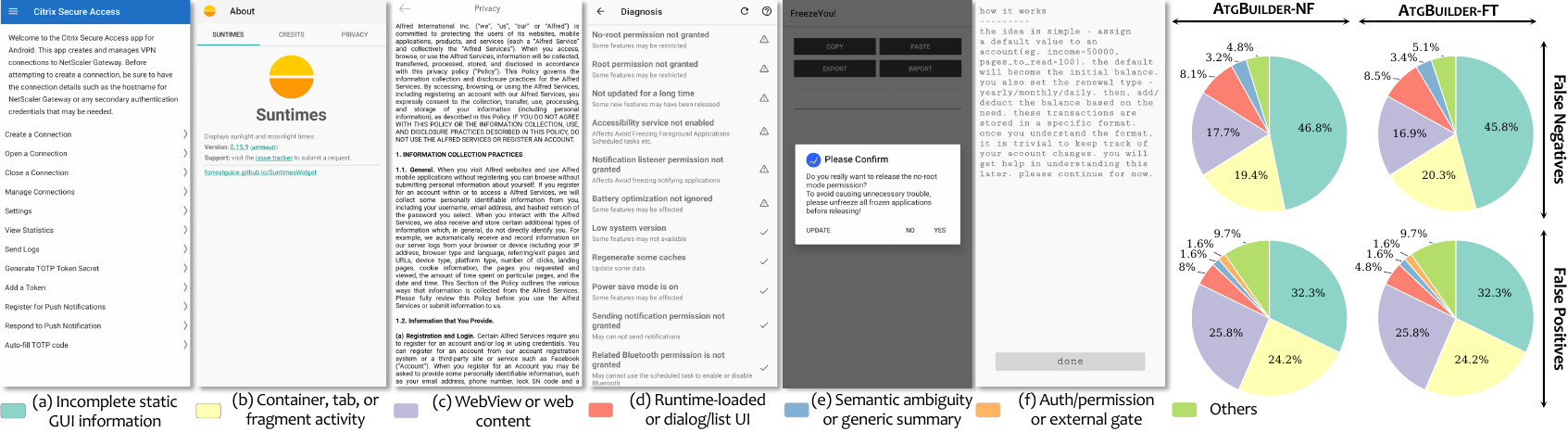}
    \caption{\revp{Examples and taxonomy distributions of \builder\/-produced false negative and false positive predictions.}}
    \label{fig:rq22_failure_example}  
\end{figure}

\revp{To better understand the limitations of \builder, Figure~\ref{fig:rq22_failure_example} summarizes false-positive and false-negative transitions in ATG construction, along with representative UI examples:
\begin{itemize}[leftmargin=2em, topsep=0pt, itemsep=0pt, parsep=0pt, partopsep=0pt]
    \item 
    \textbf{Incomplete static GUI information.}
    This is the most frequent failure source, accounting for more than 30\% of false positives and more than 45\% of false negatives.
    These failures occur when static analysis fails to capture sufficient information (such as layout and text), making the generated activity/widget summaries less informative.
    For example, in Figure~\ref{fig:rq22_failure_example}(a), the \texttt{Help} page contains many clickable help entries, but these view-tree and texts are not available in the static analysis.
    As a result, \builder\ can only rely on activity names and generic summaries, making it difficult to infer whether or not the page has different entries.

    \item
    \textbf{Container, tab, fragment, and WebView-based pages.}
    Much of the remaining failures relate to activities serving mainly as containers, where the actual page content is provided by tabs, fragments, or \texttt{WebView}s.
    This information is often configured at runtime, and is therefore difficult to recover from static evidence alone:
    The \texttt{About} page in Figure~\ref{fig:rq22_failure_example}(b), 
    for example, is organized as a tab/fragment-based container, with its concrete semantics being carried by the selected tab (not by the activity itself).
    Although the transition trigger may come from a menu or action bar, this is not explicitly available as widget evidence.
    Similarly, Figure~\ref{fig:rq22_failure_example}(c) shows a generic \texttt{WebView} page whose actual semantics depend on the runtime-provided URL or title.
    Without the concrete runtime content and source-side trigger widget, \builder\ may overestimate the relation between a settings page and a generic web-content activity.  

    \item
    \textbf{Runtime-loaded dialog/list UI.}
    Some transitions depend on widgets, dialogs, menus, or list items that are  only created during execution:
    A static analysis may only observe a generic container (such as a \texttt{ListView}), missing the concrete runtime-loaded items and their semantics.
    The diagnosis items in Figure~\ref{fig:rq22_failure_example}(d), for example, are populated after runtime checks.
    The available static evidence only provides a weak settings summary and a high-level diagnosis summary, without the concrete entry that triggers automatic diagnosis.

    \item
    \textbf{Semantic ambiguity and runtime gates.}
    These categories appear less frequently.
    Semantic ambiguity occurs when different activities have similar or overly generic summaries, making them hard to distinguish.
    The activities in Figure~\ref{fig:rq22_failure_example}(e), for example, fall into a similar settings/customization semantic space, while the static ATG lacks clear view-tree information.
    This indicates that summary-based representations are generally useful, but may still be insufficient when activity semantics are too similar.
    Runtime gates, such as authentication, permission, or external-service conditions, introduce another source of uncertainty.
    Whether the user can proceed from the instruction page in Figure~\ref{fig:rq22_failure_example}(f), depends on the onboarding and completion states. 
    Such runtime constraints are invisible to static analysis, potentially causing \builder\ to treat semantically similar onboarding activities as directly reachable.
\end{itemize}}

\revp{Fine-tuning can provide a modest recall-oriented improvement over the original model.
Compared with \builder-NF, \builder-FT increases true positives and reduces false negatives, with the number of false positives remaining unchanged.
The reduction in false negatives mainly comes from \textit{incomplete static GUI information} and \textit{WebView} cases, suggesting that fine-tuning can partially mitigate failures caused by weak static evidence and dynamic web content.}

Based on these observations, we offer the following explanations:

\textit{(1) The key difference between \builder\ and the tool baselines lies in the ATG-construction paradigm.}
The tool baselines discover transitions through runtime exploration, and produce a discrete set of observed transitions.
Such observed transitions are typically reliable once triggered (resulting in high precision).
However, this process may produce incomplete transitions under a limited time budget, and may also miss deep or condition-dependent transitions (resulting in lower recall).
In contrast, \builder\ performs link prediction over a candidate activity-pair space, and constructs plausible missing transitions based on feature similarity and graph context, thereby improving the recall.

\textit{(2) \builder\ benefits from feature-assisted representation.}
By fusing activity-semantic features and widget-trigger information, \builder\ can generalize across different naming conventions (using name or type embeddings), and UI implementations (using summary embeddings):
This enables \builder\ to predict transitions that are not easily identified by dynamic exploration.
This advantage is amplified on Android apps, where navigation often forms hub-dominated structures, and many transitions are hard to reach by some UI-context structures, such as menus and dialogs:
These structures cause dynamic tools to frequently revisit shallow states, while failing to cover deeper transitions.
\builder, in contrast, can still predict such transitions when the source and destination activities are semantically compatible.

\rqsummary{2.1}{\builder\ achieved the best effectiveness:
It recovered substantially more ground-truth transitions than tool baselines, while maintaining a good precision.
Finetuning and validation-selected thresholding ($t^*$) further yielded a better precision-recall trade-off.}

\subsubsection{Answer to RQ2.2: Robustness under Noisy Seed Supervision}
\begin{figure}[!t]
    \centering
    \includegraphics[width=\linewidth]{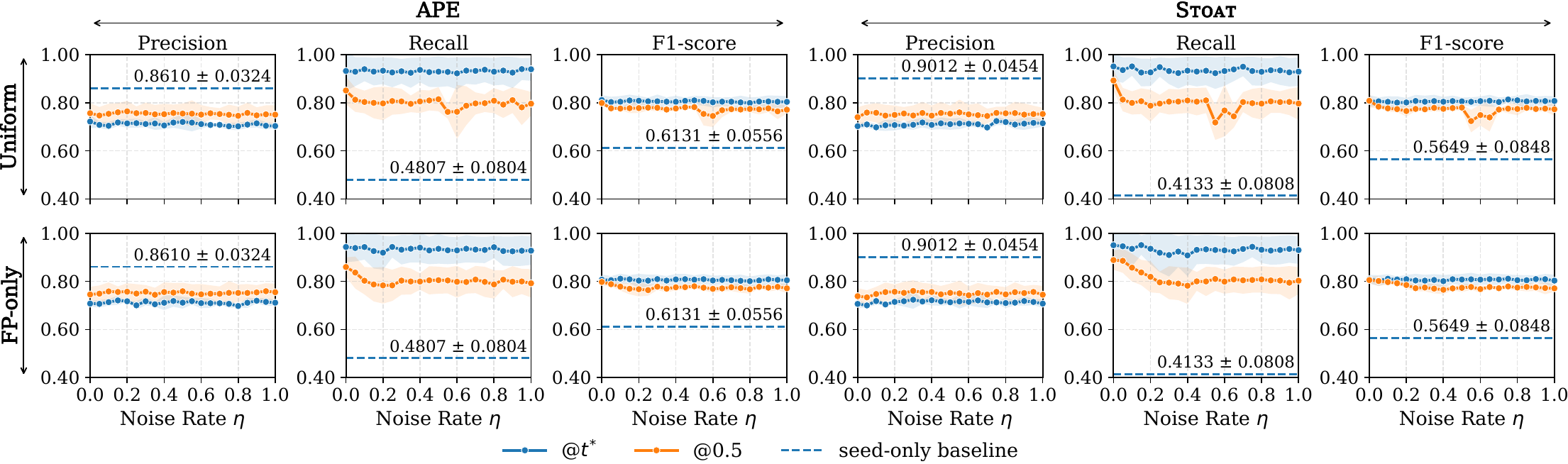}
    \caption{\revp{RQ2.2 noise-robustness curves under \textit{uniform label-flip} noise and \textit{FP-only} noise injection.}} 
    \label{fig:rq22_results}  
\end{figure}

Figure~\ref{fig:rq22_results} shows the robustness as the noise rate $\eta$ increases, using seeds derived from \ape\ and \stoat.
Based on these results, we have the following observations:
\begin{itemize}[leftmargin=2em, topsep=0pt, itemsep=0pt, parsep=0pt, partopsep=0pt]
    \item 
    \textbf{\builder\ remained robust under both noise-injection modes, and consistently outperformed the seed-only baseline.}
    Across a wide range of $\eta$, the F1-score curves remained relatively stable, and above the baseline, indicating that \builder\ did not suffer severe performance degradation, even when the supervision was substantially corrupted.

    \item 
    \textbf{Validation-selected threshold value ($@t^*$) yielded higher recall and more stable F1-score than the fixed threshold ($@0.5$).}
    Across settings, the use of $t^*$ consistently increased recall and achieved comparable (or better) F1-scores.
    The precision under $t^*$ was typically less than under $0.5$, reflecting the expected precision-recall trade-off.

    \item 
    \textbf{Uniform-flip noise affected the fixed-threshold results more noticeably than FP-only noise.}
    Under uniform flips, recall under $@0.5$ degraded at some intermediate $\eta$, whereas the $t^*$ curves remained smoother (Figure~\ref{fig:rq22_results}).
    Under FP-only injection, the F1-score remained nearly flat across $\eta$, and remained above the seed-only baseline (Figure~\ref{fig:rq22_results}).
\end{itemize}

Based on these observations, we offer the following explanations:

\textit{(1) Even under label noise, fused semantic features and graph structures still provide informative signals, enabling the model to separate transitions from non-transitions.}
Although the noise corrupts the training labels, it does not fully destroy the feature- and structural-regularities captured by \builder's fused activity and widget representations and message passing.
This allows the model to continue learning a meaningful distinction between the positive and negative edges.

\textit{(2) Validation-based thresholding adapts to score-distribution changes under noise, making $t^*$ more robust than a fixed threshold.}
A fixed threshold ($@0.5$) is more vulnerable to noisy supervision:
In contrast, $@t^*$ can adapt to the shifted score distribution by selecting a better operating point on the validation set.
This is why $@t^*$ yields consistently higher recall and more stable F1-scores across the noise rates, for both noise modes.

\rqsummary{2.2}{\builder\ remained effective when introducing noisy labels into seed supervision.
Validation-based thresholding ($t^*$) can help to select a more suitable prediction-decision threshold for ATG construction.}

\subsection{Answer to RQ3: Usefulness Evaluation}
\label{sec:res-rq3}

\begin{figure}
    \centering
    \includegraphics[width=\linewidth]{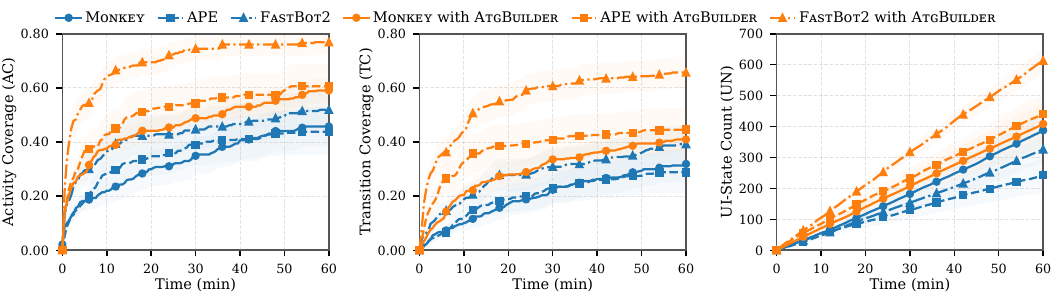}
    \caption{\revp{RQ3: Effect of \builder\ guidance on GUI-exploration effectiveness over time.}} 
    \label{fig:rq3_guided_main}  
\end{figure}

\begin{table}
\centering
\color{black}
\caption{\revp{RQ3: Endpoint effectiveness gains of \builder-guided variants over the original exploration tools.}}
\label{tab:rq3-endpoint-delta}
\scriptsize
\setlength{\tabcolsep}{3.7mm}
\begin{tabular}{lrrrrrr}
\hline
\textbf{Comparisons} & \textbf{$\Delta$AC} & \textbf{$\Delta$TC} & \textbf{$\Delta$UN} & \textbf{$\Delta$EN} & \textbf{$\Delta$ACR} & \textbf{$\Delta$UCR} \\
\hline
 \monkey\ with \textsc{AtgBuilder} over \textsc{Monkey} & +0.133 & +0.091 & +20.2 & +36.1k & -0.0005 & -0.0073 \\
 \ape\ with \textsc{AtgBuilder} over APE & +0.169 & +0.159 & +197.1 & +3.3k & +0.0140 & -0.0202 \\
 \fastbot\ with \textsc{AtgBuilder} over \textsc{FastBot2} & +0.250 & +0.265 & +286.9 & +46.6k & -0.0008 & -0.0108 \\
\hline
\end{tabular}
\end{table}

\revp{Figure~\ref{fig:rq3_guided_main} and Table~\ref{tab:rq3-endpoint-delta} present the effectiveness comparison between each original exploration tool and its \builder-guided variant.
Based on these results, we have the following observations:
\begin{itemize}[leftmargin=2em, topsep=0pt, itemsep=0pt, parsep=0pt, partopsep=0pt]
    \item 
    \textbf{\builder-guided variants improved activity and transition coverage across all three tools.}
    Compared with the original tools, the guided variants demonstrated improved activity and transition coverage.
    This was due to the predicted ATG providing activity-level navigation targets that enabled more time to be spent around under-explored activities and transitions (instead of relying only on the tool's local event-selection strategy).
    This suggests that predicted ATGs could complement existing automated GUI-exploration tools.

    \item 
    \textbf{\builder-guided variants discovered more diverse UI states.}
    As expected, the guided variants increased UI-state numbers for all three tools (with especially large gains for \ape\ and \fastbot).
    This was because reaching additional activities usually exposed new UI contexts, widgets, and screens:
    ATG guidance helped move the tools toward under-explored activity-transition directions (rather than repeatedly exercising already-explored shallow UI states).

    \item 
    \textbf{\builder\ improved exploration coverage, while preserving the original UI-event-generation behavior.}
    The guided variants achieved consistently better activity and transition coverage, and UI-State Count.
    The changes in Activity Change Rate (ACR) and UI-State Change Rate (UCR) were relatively small, about 0.02.
    This suggests that, as expected, \builder\ did not substantially change the per-event conversion behavior of the underlying tools:
    The original event-generation strategies of \monkey, \ape, and \fastbot\ were retained.
    \builder\ provided lightweight navigation guidance that directed the existing event-generation mechanisms toward more useful activity and transition spaces, leading to better performance, even when the total number of executed events also increased.
\end{itemize}}

\revp{Overall, the results show that the \builder\ complemented, not replaced, the underlying exploration engines.
The original tools generated UI events using their native strategies, with \builder\ guiding navigation toward activities and transitions that improved coverage.
This is why the benefits appeared across tools with different exploration mechanisms.}

\rqsummary{3}{\revp{\builder\ improved GUI exploration across different exploration tools under the same budget.
This suggests that \builder\ can complement existing exploration engines by providing lightweight guidance toward more activities and transitions.}}

\subsection{Threats to Validity}
\label{sec:threats-validity}

A potential threat to the validity concerns the data used to train and evaluate \builder.
We trained \builder\ on the large-scale \frontmatter~\cite{frontmatter2021} corpus, whose static-analysis-derived seeds 
can be noisy and incomplete (i.e., containing false positives and missing feasible transitions)~\cite{improvestatic2026,archidroid2023}.
Nevertheless, \frontmatter\ has been used for training and controlled comparisons in previous learning-based ATG construction~\cite{archidroid2023}.
We used this corpus for RQ1 mainly to examine ablation trends, rather than to make definitive real-world effectiveness claims.
\revp{To evaluate real-world effectiveness, we used a benchmark with manually-checked ground-truth ATGs~\cite{atgemprical2025}, where \builder\ outperformed the selected baselines.
However, \builder\ may be affected by the quality of the seed ATGs used for training, especially when they contain noisy, incomplete, or insufficiently transition labels
--- 
as revealed by the failure-case analysis in Section~\ref{sec:answer2.1}.
To further examine \builder's robustness, RQ2.2 evaluated \builder\ under synthetic label noise and dynamically-executed seed-transitions generated by \ape\ and \stoat:
The results showed that 
(1) \builder\ remained robust under noisy labels and (2) it was not tied to a specific seed-construction method, being able to use dynamically-generated seeds from other tools.}

A second threat to validity concerns the comparability of baselines and evaluation metrics.
We evaluated \builder\ and tool-based baselines against the manually-checked ground-truth ATGs~\cite{atgemprical2025}.
However, some representative learning-based methods~\cite{tacdroid2025,utg2025} were not fully comparable:
\textsc{TacDroid}~\cite{tacdroid2025} and \textsc{RLDroid}~\cite{utg2025}, for example, targeted UI-page Transition Graphs (UTGs), whose outputs were not fully aligned with our ATG prediction definition.
The source code for \textsc{ArchiDroid}~\cite{archidroid2023} is not available; therefore, we were unable to use it in our experiments.
\revp{For RQ3, we measured navigation-oriented metrics:
These capture GUI-exploration progress, but do not directly measure code coverage or fault-detection capability.}

\revp{A third threat concerns LLM-generated summaries and UI representations.
Their robustness and reproducibility may be affected by LLM configuration and stochasticity.
To reduce this threat, all apps were summarized using the same LLM and prompt template, the temperature was set to 0, and the generated summaries were cached as static artifacts.
This means that the graph-learning stage is reproducible using the saved summaries.
Another problem is that the summary length may affect the redundancy and completeness of information available to \builder:
The 30-word length is a trade-off.
A preliminary study (Appendix~\ref{sec:supp-summary-length}) showed that shorter summaries reduced completeness, and longer ones introduced noise and redundant information without consistent gains.
Furthermore, the current representation is primarily textual and metadata-based:
It does not directly encode screenshots, visual hierarchy, or visual salience.}

\revp{The fourth threat concerns the activity-level abstraction.
Activity-level ATGs cannot represent multiple UI states, such as fragments, or \texttt{WebView} content, inside the same activity.
As discussed in RQ2.1, runtime-loading components and container activities can make the available activity-level information incomplete:
Some failures reflect limitations of ATG construction, not the prediction model itself.
Nevertheless, activity-level transitions are still the basic navigation layer for Android apps, and inaccurate activity-level reachability may propagate to finer-grained graphs.
}

\section{Discussion}
\label{sec:discussions}

This section discusses the implications of our findings for the challenges in Section~\ref{sec:motivation}.

\subsection{Implications for Challenge 1: Activity Representation}

A core challenge related to this research is that similar functionality can be implemented with very different UI layouts.
This can mean that directly relying on raw layout structures may not be sufficient for ATG construction.
Our results support the design choice of prioritizing functionality summaries over UI-implementation structures.
The experiments for RQ1.4 revealed that activity-summary features contributed the largest gains, with the activity-summary-only variant achieving the best overall results. 
This indicates that LLM-generated functionality summaries can provide a more effective contribution than identifier-only representations. 
The effectiveness gains on the 98-app benchmark used in RQ2.1 suggest that these functionality-oriented activity representations generalize beyond the training corpus to real-world apps, as evidenced by the manually-checked ground-truth ATGs.

Overall, these results imply that transforming layout-related metadata into compact functionality summaries is a practical way to reduce sensitivity to layout variation and to improve cross-app generalization for ATG prediction.

\subsection{Implications for Challenge 2: Widget-Trigger Information Modeling}

Another challenge for this research is that transitions are often triggered by widgets.
Activity-pair modeling, in contrast, naturally loses some widget-trigger information. 
Another problem is that widget-trigger information extracted through static analysis can be incomplete.
Our findings suggest that widget-trigger information worked best as an auxiliary signal, rather than the main supervision source. 
Modeling widget triggers as edge attributes aligned the model input with how navigation is executed, and avoided blending a true trigger with many irrelevant within-activity widgets. 
The auxiliary widget-attribute reconstruction objective also provided an extra training signal that helped retain trigger-related cues during representation learning.
The results for RQ1.2 showed that disabling the widget-attribute reconstruction objective consistently reduced the F1-score compared with the reconstruction-enabled variants, indicating a regularization benefit.
Finally, when predicted probabilities varied across settings, choosing a validation-selected threshold offered a consistent way to determine which candidate edges should be kept for ATG construction.

In summary, our results suggest that widget-trigger information should be used as auxiliary support.
In our setup, both the seed ATG and the UI metadata are obtained through static analysis, which can introduce false positives and false negatives.
Therefore, the model should exploit widget-trigger information when a concrete transition-trigger widget is identified, but fall back to activity representations, otherwise.

\section{Conclusion}
\label{sec:conclusions}

In this paper, we have proposed \builder\ for seed-supervised ATG construction.
\builder\ fuses static UI metadata with LLM-generated activity/widget summaries, and trains a GNN-based link predictor with an auxiliary widget-attribute reconstruction objective.
\revp{Experiments have shown that \builder\ can construct high-quality ATGs, remains robust under injected seed noise, and provides useful navigation guidance for the \monkey, \ape, and \fastbot\ automated GUI-exploration tools.}
In the future, we plan to extend \builder\ in several directions:
\revp{(1) We will combine activity-level ATG prediction with runtime information to better identify dynamically loaded widgets, and hard-to-reach navigation contexts;
(2) we will enrich the current textual representation with multimodal UI features, such as screenshot-based embeddings and visual layout features, and extend \builder\ toward finer-grained UI-state transition graphs; and
(3) we will further evaluate the downstream testing utility beyond navigation coverage, including code coverage, fault detection, test-sequence generation, and GUI-oracle generation.}

 \section*{Acknowledgment}
 This work is supported by the Science and Technology Development Fund of Macau, Macao SAR, under Grant Nos. 0069/2025/RIB2 and 0021/2023/RIA1.

{
\bibliographystyle{ACM-Reference-Format}
\bibliography{major_revision}


\begin{thebibliography}{48}

%%% the Web address.

\ifx \showCODEN    \undefined \def \showCODEN     #1{\unskip}     \fi
\ifx \showDOI      \undefined \def \showDOI       #1{#1}\fi
\ifx \showISBNx    \undefined \def \showISBNx     #1{\unskip}     \fi
\ifx \showISBNxiii \undefined \def \showISBNxiii  #1{\unskip}     \fi
\ifx \showISSN     \undefined \def \showISSN      #1{\unskip}     \fi
\ifx \showLCCN     \undefined \def \showLCCN      #1{\unskip}     \fi
\ifx \shownote     \undefined \def \shownote      #1{#1}          \fi
\ifx \showarticletitle \undefined \def \showarticletitle #1{#1}   \fi
\ifx \showURL      \undefined \def \showURL       {\relax}        \fi
% invisible to TeX
\providecommand\bibfield[2]{#2}
\providecommand\bibinfo[2]{#2}
\providecommand\natexlab[1]{#1}
\providecommand\showeprint[2][]{arXiv:#2}

\bibitem[{Android Developers}(2024)]%
        {monkey}
\bibfield{author}{\bibinfo{person}{{Android Developers}}.} \bibinfo{year}{2024}\natexlab{}.
\newblock \bibinfo{title}{{UI/Application Exerciser Monkey}}.
\newblock \bibinfo{howpublished}{\url{https://developer.android.com/studio/test/other-testing-tools/monkey}}.
\newblock
\newblock
\shownote{Accessed: 2025}.


\bibitem[Baechler et~al\mbox{.}(2024)]%
        {screenai2024}
\bibfield{author}{\bibinfo{person}{Gilles Baechler}, \bibinfo{person}{Srinivas Sunkara}, \bibinfo{person}{Maria Wang}, \bibinfo{person}{Fedir Zubach}, \bibinfo{person}{Hassan Mansoor}, \bibinfo{person}{Vincent Etter}, \bibinfo{person}{Victor Carbune}, \bibinfo{person}{Jason Lin}, \bibinfo{person}{Jindong Chen}, {and} \bibinfo{person}{Abhanshu Sharma}.} \bibinfo{year}{2024}\natexlab{}.
\newblock \showarticletitle{{ScreenAI}: {A} Vision-Language Model for {UI} and Infographics Understanding}. In \bibinfo{booktitle}{\emph{Proceedings of the 33rd International Joint Conference on Artificial Intelligence (IJCAI'24)}}. \bibinfo{pages}{3058--3068}.
\newblock


\bibitem[Chen et~al\mbox{.}(2019)]%
        {storydroid2019}
\bibfield{author}{\bibinfo{person}{Sen Chen}, \bibinfo{person}{Lingling Fan}, \bibinfo{person}{Chunyang Chen}, \bibinfo{person}{Ting Su}, \bibinfo{person}{Wenhe Li}, \bibinfo{person}{Yang Liu}, {and} \bibinfo{person}{Lihua Xu}.} \bibinfo{year}{2019}\natexlab{}.
\newblock \showarticletitle{{StoryDroid}: {Automated} Generation of Storyboard for {Android} Apps}. In \bibinfo{booktitle}{\emph{Proceedings of the 41st International Conference on Software Engineering (ICSE'19)}}. \bibinfo{pages}{596--607}.
\newblock


\bibitem[Chen et~al\mbox{.}(2024)]%
        {harmonyos2024}
\bibfield{author}{\bibinfo{person}{Yige Chen}, \bibinfo{person}{Sinan Wang}, \bibinfo{person}{Yida Tao}, {and} \bibinfo{person}{Yepang Liu}.} \bibinfo{year}{2024}\natexlab{}.
\newblock \showarticletitle{Model-based {GUI} Testing for {HarmonyOS} Apps}. In \bibinfo{booktitle}{\emph{Proceedings of the 39th {IEEE/ACM} International Conference on Automated Software Engineering (ASE'24)}}. \bibinfo{pages}{2411--2414}.
\newblock


\bibitem[Dong et~al\mbox{.}(2020)]%
        {timemachine2020}
\bibfield{author}{\bibinfo{person}{Zhen Dong}, \bibinfo{person}{Marcel B{\"{o}}hme}, \bibinfo{person}{Lucia Cojocaru}, {and} \bibinfo{person}{Abhik Roychoudhury}.} \bibinfo{year}{2020}\natexlab{}.
\newblock \showarticletitle{Time-Travel Testing of Android Apps}. In \bibinfo{booktitle}{\emph{Proceedings of the 42nd International Conference on Software Engineering (ICSE'20)}}. \bibinfo{pages}{481--492}.
\newblock


\bibitem[Fan et~al\mbox{.}(2024)]%
        {screenread2024}
\bibfield{author}{\bibinfo{person}{Yue Fan}, \bibinfo{person}{Lei Ding}, \bibinfo{person}{Ching{-}Chen Kuo}, \bibinfo{person}{Shan Jiang}, \bibinfo{person}{Yang Zhao}, \bibinfo{person}{Xinze Guan}, \bibinfo{person}{Jie Yang}, \bibinfo{person}{Yi Zhang}, {and} \bibinfo{person}{Xin Wang}.} \bibinfo{year}{2024}\natexlab{}.
\newblock \showarticletitle{Read Anywhere Pointed: {L}ayout-aware {GUI} Screen Reading with Tree-of-Lens Grounding}. In \bibinfo{booktitle}{\emph{Proceedings of the 2024 Conference on Empirical Methods in Natural Language Processing (EMNLP'24)}}. \bibinfo{pages}{9503--9522}.
\newblock


\bibitem[Fazzini et~al\mbox{.}(2018)]%
        {fazzinipd2018}
\bibfield{author}{\bibinfo{person}{Mattia Fazzini}, \bibinfo{person}{Martin Prammer}, \bibinfo{person}{Marcelo d'Amorim}, {and} \bibinfo{person}{Alessandro Orso}.} \bibinfo{year}{2018}\natexlab{}.
\newblock \showarticletitle{Automatically Translating Bug Reports into Test Cases for Mobile Apps}. In \bibinfo{booktitle}{\emph{Proceedings of the 27th {ACM} {SIGSOFT} International Symposium on Software Testing and Analysis (ISSTA'18)}}. \bibinfo{pages}{141--152}.
\newblock


\bibitem[Goodfellow et~al\mbox{.}(2016)]%
        {deeplearningbook2016}
\bibfield{author}{\bibinfo{person}{Ian Goodfellow}, \bibinfo{person}{Yoshua Bengio}, {and} \bibinfo{person}{Aaron Courville}.} \bibinfo{year}{2016}\natexlab{}.
\newblock \bibinfo{booktitle}{\emph{Deep Learning}}.
\newblock \bibinfo{publisher}{MIT Press}.
\newblock
\newblock
\shownote{\url{http://www.deeplearningbook.org}}.


\bibitem[{Google}(2018)]%
        {bert_base_cased}
\bibfield{author}{\bibinfo{person}{{Google}}.} \bibinfo{year}{2018}\natexlab{}.
\newblock \bibinfo{title}{{BERT Base Model (cased)}}.
\newblock \bibinfo{howpublished}{\url{https://huggingface.co/google-bert/bert-base-cased}}.
\newblock
\newblock
\shownote{Accessed: 2026}.


\bibitem[{Google}(2026)]%
        {googleplay_categories}
\bibfield{author}{\bibinfo{person}{{Google}}.} \bibinfo{year}{2026}\natexlab{}.
\newblock \bibinfo{title}{{Android Apps on Google Play}}.
\newblock \bibinfo{howpublished}{\url{https://play.google.com/store/apps}}.
\newblock
\newblock
\shownote{Accessed: 2026}.


\bibitem[Gu et~al\mbox{.}(2019)]%
        {ape2019}
\bibfield{author}{\bibinfo{person}{Tianxiao Gu}, \bibinfo{person}{Chengnian Sun}, \bibinfo{person}{Xiaoxing Ma}, \bibinfo{person}{Chun Cao}, \bibinfo{person}{Chang Xu}, \bibinfo{person}{Yuan Yao}, \bibinfo{person}{Qirun Zhang}, \bibinfo{person}{Jian Lu}, {and} \bibinfo{person}{Zhendong Su}.} \bibinfo{year}{2019}\natexlab{}.
\newblock \showarticletitle{Practical {GUI} Testing of Android Applications via Model Abstraction and Refinement}. In \bibinfo{booktitle}{\emph{Proceedings of the 41st International Conference on Software Engineering (ICSE'19)}}. \bibinfo{pages}{269--280}.
\newblock


\bibitem[Guo et~al\mbox{.}(2025)]%
        {utg2025}
\bibfield{author}{\bibinfo{person}{Wunan Guo}, \bibinfo{person}{Zhen Dong}, \bibinfo{person}{Liwei Shen}, \bibinfo{person}{Daihong Zhou}, \bibinfo{person}{Bin Hu}, \bibinfo{person}{Chen Zhang}, {and} \bibinfo{person}{Hai Xue}.} \bibinfo{year}{2025}\natexlab{}.
\newblock \showarticletitle{Effectively Modeling {UI} Transition Graphs for {Android} Apps Via Reinforcement Learning}. In \bibinfo{booktitle}{\emph{Proceedings of the 33rd {IEEE/ACM} International Conference on Program Comprehension (ICPC'25)}}. \bibinfo{pages}{13--24}.
\newblock


\bibitem[He et~al\mbox{.}(2025)]%
        {summary2025}
\bibfield{author}{\bibinfo{person}{Yiling He}, \bibinfo{person}{Hongyu She}, \bibinfo{person}{Xingzhi Qian}, \bibinfo{person}{Xinran Zheng}, \bibinfo{person}{Zhuo Chen}, \bibinfo{person}{Zhan Qin}, {and} \bibinfo{person}{Lorenzo Cavallaro}.} \bibinfo{year}{2025}\natexlab{}.
\newblock \showarticletitle{On Benchmarking Code LLMs for Android Malware Analysis}. In \bibinfo{booktitle}{\emph{Proceedings of the 34th ACM SIGSOFT International Symposium on Software Testing and Analysis (ISSTA Companion'25)}}. \bibinfo{pages}{153--160}.
\newblock


\bibitem[Kipf and Welling(2017)]%
        {gcn2017}
\bibfield{author}{\bibinfo{person}{Thomas~N. Kipf} {and} \bibinfo{person}{Max Welling}.} \bibinfo{year}{2017}\natexlab{}.
\newblock \showarticletitle{Semi-Supervised Classification with Graph Convolutional Networks}. In \bibinfo{booktitle}{\emph{Proceedings of the 5th International Conference on Learning Representations (ICLR'17)}}.
\newblock


\bibitem[Kong et~al\mbox{.}(2018)]%
        {mobiletestingsurvey2018}
\bibfield{author}{\bibinfo{person}{Pingfan Kong}, \bibinfo{person}{Li Li}, \bibinfo{person}{Jun Gao}, \bibinfo{person}{Kui Liu}, \bibinfo{person}{Tegawend{\'e}~F Bissyand{\'e}}, {and} \bibinfo{person}{Jacques Klein}.} \bibinfo{year}{2018}\natexlab{}.
\newblock \showarticletitle{Automated Testing of {A}ndroid Apps: {A} Systematic Literature Review}.
\newblock \bibinfo{journal}{\emph{IEEE Transactions on Reliability}} \bibinfo{volume}{68}, \bibinfo{number}{1} (\bibinfo{year}{2018}), \bibinfo{pages}{45--66}.
\newblock


\bibitem[Kuznetsov et~al\mbox{.}(2018)]%
        {KuznetsovAGZ18}
\bibfield{author}{\bibinfo{person}{Konstantin Kuznetsov}, \bibinfo{person}{Vitalii Avdiienko}, \bibinfo{person}{Alessandra Gorla}, {and} \bibinfo{person}{Andreas Zeller}.} \bibinfo{year}{2018}\natexlab{}.
\newblock \showarticletitle{Analyzing the user interface of Android apps}. In \bibinfo{booktitle}{\emph{Proceedings of the 5th International Conference on Mobile Software Engineering and Systems (MOBILESoft@ICSE'18)}}. \bibinfo{pages}{84--87}.
\newblock


\bibitem[Kuznetsov et~al\mbox{.}(2021)]%
        {frontmatter2021}
\bibfield{author}{\bibinfo{person}{Konstantin Kuznetsov}, \bibinfo{person}{Chen Fu}, \bibinfo{person}{Song Gao}, \bibinfo{person}{David~N. Jansen}, \bibinfo{person}{Lijun Zhang}, {and} \bibinfo{person}{Andreas Zeller}.} \bibinfo{year}{2021}\natexlab{}.
\newblock \showarticletitle{Frontmatter: {M}ining {Android} User Interfaces at Scale}. In \bibinfo{booktitle}{\emph{Proceedings of the 29th {ACM} Joint European Software Engineering Conference and Symposium on the Foundations of Software Engineering (ESEC/FSE'21)}}. \bibinfo{pages}{1580--1584}.
\newblock


\bibitem[Lai and Rubin(2019)]%
        {julia_rubin_goal_explorer}
\bibfield{author}{\bibinfo{person}{Duling Lai} {and} \bibinfo{person}{Julia Rubin}.} \bibinfo{year}{2019}\natexlab{}.
\newblock \showarticletitle{Goal-Driven Exploration for Android Applications}. In \bibinfo{booktitle}{\emph{Proceedings of the 34th {IEEE/ACM} International Conference on Automated Software Engineering (ASE'19)}}. \bibinfo{pages}{115--127}.
\newblock


\bibitem[Laine and Aila(2017)]%
        {seedconsistency2017}
\bibfield{author}{\bibinfo{person}{Samuli Laine} {and} \bibinfo{person}{Timo Aila}.} \bibinfo{year}{2017}\natexlab{}.
\newblock \showarticletitle{Temporal Ensembling for Semi-Supervised Learning}. In \bibinfo{booktitle}{\emph{Proceedings of the 5th International Conference on Learning Representations (ICLR'17)}}.
\newblock


\bibitem[Li et~al\mbox{.}(2018)]%
        {oversmoothing2018}
\bibfield{author}{\bibinfo{person}{Qimai Li}, \bibinfo{person}{Zhichao Han}, {and} \bibinfo{person}{Xiao{-}Ming Wu}.} \bibinfo{year}{2018}\natexlab{}.
\newblock \showarticletitle{Deeper Insights Into Graph Convolutional Networks for Semi-Supervised Learning}. In \bibinfo{booktitle}{\emph{Proceedings of the 32nd {AAAI} Conference on Artificial Intelligence (AAAI'18)}}. \bibinfo{pages}{3538--3545}.
\newblock


\bibitem[Li et~al\mbox{.}(2019)]%
        {humanoid2019}
\bibfield{author}{\bibinfo{person}{Yuanchun Li}, \bibinfo{person}{Ziyue Yang}, \bibinfo{person}{Yao Guo}, {and} \bibinfo{person}{Xiangqun Chen}.} \bibinfo{year}{2019}\natexlab{}.
\newblock \showarticletitle{Humanoid: {A} Deep Learning-Based Approach to Automated Black-Box Android App Testing}. In \bibinfo{booktitle}{\emph{Proceedings of the 34th {IEEE/ACM} International Conference on Automated Software Engineering (ASE'19)}}. \bibinfo{pages}{1070--1073}.
\newblock


\bibitem[{LibreTube contributors}(2021)]%
        {libretube}
\bibfield{author}{\bibinfo{person}{{LibreTube contributors}}.} \bibinfo{year}{2021}\natexlab{}.
\newblock \bibinfo{title}{LibreTube}.
\newblock \bibinfo{howpublished}{\url{https://github.com/libre-tube/LibreTube}}.
\newblock
\newblock
\shownote{Accessed: 2026}.


\bibitem[Liu et~al\mbox{.}(2025)]%
        {atgemprical2025}
\bibfield{author}{\bibinfo{person}{Jiakun Liu}, \bibinfo{person}{Peixin Zhang}, \bibinfo{person}{Han Hu}, \bibinfo{person}{Yonghui Liu}, \bibinfo{person}{Wei Minn}, \bibinfo{person}{Ferdian Thung}, \bibinfo{person}{Shahar Maoz}, \bibinfo{person}{Eran Toch}, \bibinfo{person}{Debin Gao}, {and} \bibinfo{person}{David Lo}.} \bibinfo{year}{2025}\natexlab{}.
\newblock \showarticletitle{Activity Transition Graph Generation: {H}ow Far Are We?}
\newblock \bibinfo{journal}{\emph{ACM Transactions on Software Engineering and Methodology}} (\bibinfo{year}{2025}).
\newblock
\urldef\tempurl%
\url{https://doi.org/10.1145/3776553}
\showDOI{\tempurl}
\newblock
\shownote{Just Accepted}.


\bibitem[Liu et~al\mbox{.}(2023a)]%
        {fillintheblank2023}
\bibfield{author}{\bibinfo{person}{Zhe Liu}, \bibinfo{person}{Chunyang Chen}, \bibinfo{person}{Junjie Wang}, \bibinfo{person}{Xing Che}, \bibinfo{person}{Yuekai Huang}, \bibinfo{person}{Jun Hu}, {and} \bibinfo{person}{Qing Wang}.} \bibinfo{year}{2023}\natexlab{a}.
\newblock \showarticletitle{{F}ill in the Blank: {C}ontext-aware Automated Text Input Generation for Mobile {GUI} Testing}. In \bibinfo{booktitle}{\emph{Proceedings of the 45th International Conference on Software Engineering}}. \bibinfo{pages}{1355--1367}.
\newblock


\bibitem[Liu et~al\mbox{.}(2023b)]%
        {archidroid2023}
\bibfield{author}{\bibinfo{person}{Zhe Liu}, \bibinfo{person}{Chunyang Chen}, \bibinfo{person}{Junjie Wang}, \bibinfo{person}{Yuhui Su}, \bibinfo{person}{Yuekai Huang}, \bibinfo{person}{Jun Hu}, {and} \bibinfo{person}{Qing Wang}.} \bibinfo{year}{2023}\natexlab{b}.
\newblock \showarticletitle{{Ex pede Herculem}: {A}ugmenting {Activity} Transition Graph for Apps via Graph Convolution Network}. In \bibinfo{booktitle}{\emph{Proceedings of the 45th {IEEE/ACM} International Conference on Software Engineering (ICSE'23)}}. \bibinfo{pages}{1983--1995}.
\newblock


\bibitem[Lu et~al\mbox{.}(2025)]%
        {tacdroid2025}
\bibfield{author}{\bibinfo{person}{Yanchen Lu}, \bibinfo{person}{Hongyu Lin}, \bibinfo{person}{Zehua He}, \bibinfo{person}{Haitao Xu}, \bibinfo{person}{Zhao Li}, \bibinfo{person}{Shuai Hao}, \bibinfo{person}{Liu Wang}, \bibinfo{person}{Haoyu Wang}, {and} \bibinfo{person}{Kui Ren}.} \bibinfo{year}{2025}\natexlab{}.
\newblock \showarticletitle{{TacDroid}: {D}etection of Illicit Apps Through Hybrid Analysis of {UI}-Based Transition Graphs}. In \bibinfo{booktitle}{\emph{Proceedings of the 47th {IEEE/ACM} International Conference on Software Engineering (ICSE'25)}}. \bibinfo{pages}{2790--2802}.
\newblock


\bibitem[Lv et~al\mbox{.}(2022)]%
        {fastbot22022}
\bibfield{author}{\bibinfo{person}{Zhengwei Lv}, \bibinfo{person}{Chao Peng}, \bibinfo{person}{Zhao Zhang}, \bibinfo{person}{Ting Su}, \bibinfo{person}{Kai Liu}, {and} \bibinfo{person}{Ping Yang}.} \bibinfo{year}{2022}\natexlab{}.
\newblock \showarticletitle{Fastbot2: {R}eusable Automated Model-based {GUI} Testing for {Android} Enhanced by Reinforcement Learning}. In \bibinfo{booktitle}{\emph{Proceedings of the 37th {IEEE/ACM} International Conference on Automated Software Engineering (ASE'22)}}. \bibinfo{pages}{135:1--135:5}.
\newblock


\bibitem[Nguyen et~al\mbox{.}(2014)]%
        {guitar2014}
\bibfield{author}{\bibinfo{person}{Bao~N. Nguyen}, \bibinfo{person}{Bryan Robbins}, \bibinfo{person}{Ishan Banerjee}, {and} \bibinfo{person}{Atif~M. Memon}.} \bibinfo{year}{2014}\natexlab{}.
\newblock \showarticletitle{{GUITAR:} An Innovative Tool for Automated Testing of {GUI}-driven Software}.
\newblock \bibinfo{journal}{\emph{Automated Software Engineering}} \bibinfo{volume}{21}, \bibinfo{number}{1} (\bibinfo{year}{2014}), \bibinfo{pages}{65--105}.
\newblock


\bibitem[Oono and Suzuki(2020)]%
        {oversmoothing2020}
\bibfield{author}{\bibinfo{person}{Kenta Oono} {and} \bibinfo{person}{Taiji Suzuki}.} \bibinfo{year}{2020}\natexlab{}.
\newblock \showarticletitle{Graph Neural Networks Exponentially Lose Expressive Power for Node Classification}. In \bibinfo{booktitle}{\emph{Proceedings of the 8th International Conference on Learning Representations (ICLR'20)}}.
\newblock


\bibitem[{OpenAI}(2024)]%
        {gpt4o}
\bibfield{author}{\bibinfo{person}{{OpenAI}}.} \bibinfo{year}{2024}\natexlab{}.
\newblock \bibinfo{title}{{GPT-4o}}.
\newblock \bibinfo{howpublished}{\url{https://openai.com/index/hello-gpt-4o}}.
\newblock
\newblock
\shownote{Accessed: 2025}.


\bibitem[Pan et~al\mbox{.}(2020)]%
        {qtesting2020}
\bibfield{author}{\bibinfo{person}{Minxue Pan}, \bibinfo{person}{An Huang}, \bibinfo{person}{Guoxin Wang}, \bibinfo{person}{Tian Zhang}, {and} \bibinfo{person}{Xuandong Li}.} \bibinfo{year}{2020}\natexlab{}.
\newblock \showarticletitle{Reinforcement Learning Based Curiosity-Driven Testing of {Android} Applications}. In \bibinfo{booktitle}{\emph{Proceedings of the 29th {ACM} {SIGSOFT} International Symposium on Software Testing and Analysis (ISSTA'20)}}. \bibinfo{pages}{153--164}.
\newblock


\bibitem[Reed et~al\mbox{.}(2015)]%
        {bootstrapping2015}
\bibfield{author}{\bibinfo{person}{Scott~E. Reed}, \bibinfo{person}{Honglak Lee}, \bibinfo{person}{Dragomir Anguelov}, \bibinfo{person}{Christian Szegedy}, \bibinfo{person}{Dumitru Erhan}, {and} \bibinfo{person}{Andrew Rabinovich}.} \bibinfo{year}{2015}\natexlab{}.
\newblock \showarticletitle{Training Deep Neural Networks on Noisy Labels with Bootstrapping}. In \bibinfo{booktitle}{\emph{Proceedings of the 3rd International Conference on Learning Representations (ICLR'15)}}.
\newblock


\bibitem[Reimers and Gurevych(2019)]%
        {sentencebert2019}
\bibfield{author}{\bibinfo{person}{Nils Reimers} {and} \bibinfo{person}{Iryna Gurevych}.} \bibinfo{year}{2019}\natexlab{}.
\newblock \showarticletitle{{Sentence-BERT}: {S}entence Embeddings using Siamese {BERT}-Networks}. In \bibinfo{booktitle}{\emph{Proceedings of the 2019 Conference on Empirical Methods in Natural Language Processing and the 9th International Joint Conference on Natural Language Processing (EMNLP-IJCNLP'19)}}. \bibinfo{pages}{3980--3990}.
\newblock


\bibitem[Samhi et~al\mbox{.}(2026)]%
        {improvestatic2026}
\bibfield{author}{\bibinfo{person}{Jordan Samhi}, \bibinfo{person}{Ren\'{e} Just}, \bibinfo{person}{Michael~D. Ernst}, \bibinfo{person}{Tegawend\'{e}~F. Bissyand\'{e}}, {and} \bibinfo{person}{Jacques Klein}.} \bibinfo{year}{2026}\natexlab{}.
\newblock \showarticletitle{Resolving Conditional Implicit Calls to Improve Static and Dynamic Analysis in Android Apps}.
\newblock \bibinfo{journal}{\emph{ACM Transactions on Software Engineering and Methodology}} \bibinfo{volume}{35}, \bibinfo{number}{2} (\bibinfo{year}{2026}).
\newblock


\bibitem[{Sentence-BERT}(2020)]%
        {all_minilm_l12_v2}
\bibfield{author}{\bibinfo{person}{{Sentence-BERT}}.} \bibinfo{year}{2020}\natexlab{}.
\newblock \bibinfo{title}{{all-MiniLM-L12-v2}}.
\newblock \bibinfo{howpublished}{\url{https://huggingface.co/sentence-transformers/all-MiniLM-L12-v2}}.
\newblock
\newblock
\shownote{Accessed: 2026}.


\bibitem[{Statista}(2025)]%
        {android202511}
\bibfield{author}{\bibinfo{person}{{Statista}}.} \bibinfo{year}{2025}\natexlab{}.
\newblock \bibinfo{title}{Market Share of Mobile Operating Systems Worldwide from 2009 to 2025, by Quarter}.
\newblock \bibinfo{howpublished}{\url{https://www.statista.com/statistics/272698/global-market-share-held-by-mobile-operating-systems-since-2009/}}.
\newblock
\newblock
\shownote{Accessed: 2026}.


\bibitem[Su et~al\mbox{.}(2017)]%
        {stoat2017}
\bibfield{author}{\bibinfo{person}{Ting Su}, \bibinfo{person}{Guozhu Meng}, \bibinfo{person}{Yuting Chen}, \bibinfo{person}{Ke Wu}, \bibinfo{person}{Weiming Yang}, \bibinfo{person}{Yao Yao}, \bibinfo{person}{Geguang Pu}, \bibinfo{person}{Yang Liu}, {and} \bibinfo{person}{Zhendong Su}.} \bibinfo{year}{2017}\natexlab{}.
\newblock \showarticletitle{Guided, Stochastic Model-based {GUI} Testing of {Android} Apps}. In \bibinfo{booktitle}{\emph{Proceedings of the 2017 11th Joint Meeting on Foundations of Software Engineering (ESEC/FSE'17)}}. \bibinfo{pages}{245--256}.
\newblock


\bibitem[{Team NewPipe}(2015)]%
        {newpipe}
\bibfield{author}{\bibinfo{person}{{Team NewPipe}}.} \bibinfo{year}{2015}\natexlab{}.
\newblock \bibinfo{title}{NewPipe}.
\newblock \bibinfo{howpublished}{\url{https://github.com/TeamNewPipe/NewPipe}}.
\newblock
\newblock
\shownote{Accessed: 2026}.


\bibitem[Wang et~al\mbox{.}(2021)]%
        {toller2021}
\bibfield{author}{\bibinfo{person}{Wenyu Wang}, \bibinfo{person}{Wing Lam}, {and} \bibinfo{person}{Tao Xie}.} \bibinfo{year}{2021}\natexlab{}.
\newblock \showarticletitle{An Infrastructure Approach to Improving Effectiveness of {Android} {UI} Testing Tools}. In \bibinfo{booktitle}{\emph{Proceedings of the 30th {ACM} {SIGSOFT} International Symposium on Software Testing and Analysis (ISSTA'21)}}. \bibinfo{publisher}{{ACM}}, \bibinfo{pages}{165--176}.
\newblock


\bibitem[Wu et~al\mbox{.}(2021)]%
        {gnnsurvey2021}
\bibfield{author}{\bibinfo{person}{Zonghan Wu}, \bibinfo{person}{Shirui Pan}, \bibinfo{person}{Fengwen Chen}, \bibinfo{person}{Guodong Long}, \bibinfo{person}{Chengqi Zhang}, {and} \bibinfo{person}{Philip~S. Yu}.} \bibinfo{year}{2021}\natexlab{}.
\newblock \showarticletitle{A Comprehensive Survey on Graph Neural Networks}.
\newblock \bibinfo{journal}{\emph{IEEE Transactions on Neural Networks and Learning Systems}} \bibinfo{volume}{32}, \bibinfo{number}{1} (\bibinfo{year}{2021}), \bibinfo{pages}{4--24}.
\newblock


\bibitem[Xiao et~al\mbox{.}(2019)]%
        {iconintent2019}
\bibfield{author}{\bibinfo{person}{Xusheng Xiao}, \bibinfo{person}{Xiaoyin Wang}, \bibinfo{person}{Zhihao Cao}, \bibinfo{person}{Hanlin Wang}, {and} \bibinfo{person}{Peng Gao}.} \bibinfo{year}{2019}\natexlab{}.
\newblock \showarticletitle{{IconIntent}: {A}utomatic Identification of Sensitive {UI} Widgets based on Icon Classification for {Android} Apps}. In \bibinfo{booktitle}{\emph{Proceedings of the 41st International Conference on Software Engineering (ICSE'19)}}. \bibinfo{pages}{257--268}.
\newblock


\bibitem[Xu et~al\mbox{.}(2019)]%
        {gin2019}
\bibfield{author}{\bibinfo{person}{Keyulu Xu}, \bibinfo{person}{Weihua Hu}, \bibinfo{person}{Jure Leskovec}, {and} \bibinfo{person}{Stefanie Jegelka}.} \bibinfo{year}{2019}\natexlab{}.
\newblock \showarticletitle{How Powerful are Graph Neural Networks?}. In \bibinfo{booktitle}{\emph{Proceedings of the 7th International Conference on Learning Representations (ICLR'19)}}.
\newblock


\bibitem[Yang et~al\mbox{.}(2018)]%
        {staticwtg2018}
\bibfield{author}{\bibinfo{person}{Shengqian Yang}, \bibinfo{person}{Haowei Wu}, \bibinfo{person}{Hailong Zhang}, \bibinfo{person}{Yan Wang}, \bibinfo{person}{Chandrasekar Swaminathan}, \bibinfo{person}{Dacong Yan}, {and} \bibinfo{person}{Atanas Rountev}.} \bibinfo{year}{2018}\natexlab{}.
\newblock \showarticletitle{Static window transition graphs for {Android}}.
\newblock \bibinfo{journal}{\emph{Automated Software Engineering}} \bibinfo{volume}{25}, \bibinfo{number}{4} (\bibinfo{year}{2018}), \bibinfo{pages}{833--873}.
\newblock


\bibitem[Yang et~al\mbox{.}(2022)]%
        {permdroid2022}
\bibfield{author}{\bibinfo{person}{Shuaihao Yang}, \bibinfo{person}{Zigang Zeng}, {and} \bibinfo{person}{Wei Song}.} \bibinfo{year}{2022}\natexlab{}.
\newblock \showarticletitle{PermDroid: {A}utomatically Testing Permission-Related Behaviour of {Android} Applications}. In \bibinfo{booktitle}{\emph{Proceedings of the 31st {ACM} {SIGSOFT} International Symposium on Software Testing and Analysis (ISSTA'22)}}. \bibinfo{pages}{593--604}.
\newblock


\bibitem[{Zenodo}(2021)]%
        {frontmatter_dataset}
\bibfield{author}{\bibinfo{person}{{Zenodo}}.} \bibinfo{year}{2021}\natexlab{}.
\newblock \bibinfo{title}{{Frontmatter Dataset}}.
\newblock \bibinfo{howpublished}{\url{https://zenodo.org/records/5084655}}.
\newblock
\newblock
\shownote{Accessed: 2026}.


\bibitem[Zhang and Chen(2018)]%
        {linkprediction2018}
\bibfield{author}{\bibinfo{person}{Muhan Zhang} {and} \bibinfo{person}{Yixin Chen}.} \bibinfo{year}{2018}\natexlab{}.
\newblock \showarticletitle{Link Prediction Based on Graph Neural Networks}. In \bibinfo{booktitle}{\emph{Proceedings of the 31st Annual Conference on Neural Information Processing Systems (NeurIPS'18)}}. \bibinfo{pages}{5171--5181}.
\newblock


\bibitem[Zhang et~al\mbox{.}(2023)]%
        {scenedroid2023}
\bibfield{author}{\bibinfo{person}{Xiangyu Zhang}, \bibinfo{person}{Lingling Fan}, \bibinfo{person}{Sen Chen}, \bibinfo{person}{Yucheng Su}, {and} \bibinfo{person}{Boyuan Li}.} \bibinfo{year}{2023}\natexlab{}.
\newblock \showarticletitle{Scene-Driven Exploration and {GUI} Modeling for {Android} Apps}. In \bibinfo{booktitle}{\emph{Proceedings of the 38th {IEEE/ACM} International Conference on Automated Software Engineering (ASE'23)}}. \bibinfo{pages}{1251--1262}.
\newblock


\bibitem[Zhao et~al\mbox{.}(2018)]%
        {zhaoyixue2018}
\bibfield{author}{\bibinfo{person}{Yixue Zhao}, \bibinfo{person}{Marcelo~Schmitt Laser}, \bibinfo{person}{Yingjun Lyu}, {and} \bibinfo{person}{Nenad Medvidovic}.} \bibinfo{year}{2018}\natexlab{}.
\newblock \showarticletitle{Leveraging Program Analysis to Reduce User-Perceived Latency in Mobile Applications}. In \bibinfo{booktitle}{\emph{Proceedings of the 40th International Conference on Software Engineering (ICSE'18)}}. \bibinfo{pages}{176--186}.
\newblock


\end{thebibliography}
}

\appendix
\clearpage

\newpage
\setcounter{page}{1} 
\pagenumbering{arabic} 
\section*{Appendices} 
\addcontentsline{toc}{section}{Appendices} 
\label{asection}

\setcounter{table}{0}
\renewcommand{\thetable}{\thesection.\arabic{table}}

This document provides supplementary details referenced in the main paper, including:
(1) the full LLM prompt template for summary generation;
(2) the summary-length selection procedure; and
(3) additional ablation details on the GCN vs.\ GIN comparison.

\section{Full Prompt Template for LLM-based Summary Generation}
\label{sec:supp-prompt}

This section provides the exact prompt templates used to generate one-sentence summaries of functionality for activities and widgets.
In the templates below, placeholders (e.g., \textttsplie{[MASK\_ID]}, \textttsplie{[MASK\_STRUCTURE]}) are instantiated with the corresponding metadata extracted for each activity/widget.

\begin{lstlisting}[basicstyle=\scriptsize\ttfamily, breaklines=true]
activity_prompt_template: |
  As an Android app tester, given an Activity's info:
  - Activity id: [MASK_ID]
  - Activity name: [MASK_NAME]
  - Detailed structure: [MASK_STRUCTURE]
  
  Succinctly and accurately summarize the core purpose of this Activity in one sentence.
  Your Response must follow the requirements:
  1. In English and in one concise sentence (<= 30 English words)
  2. Accurately reflects the main function/purpose
  3. No extra info/explanations
  4. Return as a pure JSON object wrapped in triple backticks (code block markers):
  ```{
    "activity_id": "[MASK_ID]",
    "activity_name": "[MASK_NAME]",
    "purpose": "YOUR_ANSWER"
  }```

retry_prompt_template: |
  The previous response did not meet the requirements. Please try again to summarize the core purpose of the Activity in one sentence, following the specified format and constraints:
  - Activity id: [MASK_ID]
  - Activity name: [MASK_NAME]
  - Detailed structure: [MASK_STRUCTURE]
  
  Succinctly and accurately summarize the core purpose of this Activity in one sentence.
  Your Response must follow the requirements:
  1. In English and in one concise sentence (<= 30 English words)
  2. Accurately reflects the main function/purpose
  3. No extra info/explanations
  4. Return as a pure JSON object wrapped in triple backticks (code block markers):
  ```{
    "activity_id": "[MASK_ID]",
    "activity_name": "[MASK_NAME]",
    "purpose": "YOUR_ANSWER"
  }```

widget_prompt_template: |
  As an Android app tester, given a Widget's info:
  - Widget id: [MASK_ID]
  - Widget type: [MASK_TYPE]
  - Widget structure: [MASK_CONTENT]
  
  Succinctly and accurately summarize the core purpose/function of this Widget in one sentence.
  Your Response must follow the requirements:
  1. In English and in one concise sentence (<= 30 English words)
  2. Accurately reflects the main function/purpose
  3. No extra info/explanations
  4. Return as a pure JSON object wrapped in triple backticks (code block markers):
  ```{
    "widget_id": "[MASK_ID]",
    "widget_type": "[MASK_TYPE]",
    "purpose": "YOUR_ANSWER"
  }```

retry_widget_prompt_template: |
  The previous response did not meet the requirements. Please try again to summarize the core purpose of the Activity in one sentence, following the specified format and constraints:
  - Widget id: [MASK_ID]
  - Widget type: [MASK_TYPE]
  - Widget structure: [MASK_CONTENT]
  
  Succinctly and accurately summarize the core purpose/function of this Widget in one sentence.
  Your Response must follow the requirements:
  1. In English and in one concise sentence (<= 30 English words)
  2. Accurately reflects the main function/purpose
  3. No extra info/explanations
  4. Return as a pure JSON object wrapped in triple backticks (code block markers):
  ```{
    "widget_id": "[MASK_ID]",
    "widget_type": "[MASK_TYPE]",
    "purpose": "YOUR_ANSWER"
  }```
\end{lstlisting}

These prompts are used consistently across apps to encourage concise, consistent summaries for subsequent embedding and feature fusion.

\section{Summary Length Selection}
\label{sec:supp-summary-length}

\builder\ uses one-sentence LLM-generated summaries to represent the functionality of activities and widgets.
We used a 30-word upper bound in the main experiments.
To examine this design choice, we conducted a summary-level sensitivity analysis by regenerating summaries under word budgets from 10 to 100 words, in increments of 10 words.
This analysis was designed to isolate the effect of summary length: we used the same LLM (\textsc{GPT-4o}), the same prompt templates in Section~\ref{sec:supp-prompt}, the same input evidence, a fixed temperature of 0, and the same JSON output constraint.
Only the word-budget requirement in the prompt was changed.

The experiment sampled 100 apps from the training-app pool, excluding the ground-truth evaluation apps used in the main ATG experiments.
In total, the sampled set contained 976 summarized entities, including 695 activities and 281 widgets.
For activity summaries, the input evidence was the full extracted activity structure.
For widget summaries, the input evidence was the structure of trigger widgets that appeared in transitions.
The purpose of this analysis was not to retrain the graph model for every word budget, but to evaluate whether a given budget preserved useful UI evidence while avoiding unsupported or redundant text.

We used the following metrics.
Completeness was measured by \textit{evidence-term recall}, i.e., the fraction of key terms from the raw UI evidence that were covered by the generated summary, and \textit{action-object coverage}, i.e., the coverage of action/object cues such as \textit{add}, \textit{edit}, \textit{delete}, \textit{settings}, \textit{export}, and \textit{login}.
The evidence terms were extracted from activity names, widget types, resource identifiers, visible text, content descriptions, UI classes, and action-related terms.
Noise was measured by \textit{unsupported-term rate}, the fraction of summary terms not supported by the raw UI evidence, \textit{generic verbosity}, the fraction of generic UI description terms such as \textit{screen}, \textit{interface}, \textit{allows}, \textit{provides}, and \textit{information}, and \textit{redundancy}, the rate of repeated words or low-information repeated phrases.
We also recorded the provider-side token count and LLM latency.
Finally, as a lightweight transition-level proxy without retraining the graph model, we used the source-activity summary and trigger-widget summary to retrieve the true destination activity among candidate destinations from the same app, and reported Hit@3 and MRR.

\setcounter{table}{0}
                \renewcommand{\thetable}{A.\arabic{table}}  
\begin{table*}[!t]
    \centering
    \scriptsize
    \setlength{\tabcolsep}{1.7mm}
    \renewcommand{\arraystretch}{1.02}
    \captionsetup{skip=2pt}
    \caption{Summary-length sensitivity results for representative word budgets.}
    \label{tab:supp-summary-length-metrics}
    \resizebox{\textwidth}{!}{%
    \begin{tabular}{@{}c c r r r r r r r r@{}}
        \toprule
        \textbf{Type} &
        \textbf{Budget} &
        \textbf{N} &
        \textbf{Avg. words} &
        \textbf{Avg. tokens} &
        \textbf{Latency (ms)} &
        \textbf{Evidence recall} &
        \textbf{Action-object cov.} &
        \textbf{Unsupported rate} &
        \textbf{Redundancy} \\
        \midrule
        Activity & 10  & 695 & 6.70  & 1579.02 & 3664.52 & 0.1149 & 0.1053 & 0.7039 & 0.0004 \\
        Activity & 20  & 695 & 10.99 & 1584.35 & 3795.54 & 0.1463 & 0.1386 & 0.7158 & 0.0029 \\
        Activity & 30  & 695 & 12.13 & 1585.83 & 3899.66 & 0.1443 & 0.1454 & 0.7265 & 0.0036 \\
        Activity & 40  & 695 & 13.14 & 1587.09 & 3826.32 & 0.1446 & 0.1668 & 0.7363 & 0.0038 \\
        Activity & 60  & 695 & 14.67 & 1589.12 & 3856.01 & 0.1541 & 0.1764 & 0.7313 & 0.0072 \\
        Activity & 100 & 695 & 15.73 & 1590.16 & 3938.79 & 0.1516 & 0.1657 & 0.7453 & 0.0079 \\
        \midrule
        Widget & 10  & 281 & 6.24  & 354.51 & 3488.29 & 0.1959 & 0.3049 & 0.6079 & 0.0000 \\
        Widget & 20  & 281 & 9.83  & 359.05 & 3538.39 & 0.2265 & 0.4098 & 0.6832 & 0.0056 \\
        Widget & 30  & 281 & 10.78 & 360.44 & 3636.57 & 0.2446 & 0.4116 & 0.6775 & 0.0076 \\
        Widget & 40  & 281 & 11.96 & 361.76 & 3659.02 & 0.2630 & 0.4380 & 0.6807 & 0.0116 \\
        Widget & 60  & 281 & 13.37 & 363.56 & 3691.12 & 0.2952 & 0.4425 & 0.6765 & 0.0154 \\
        Widget & 100 & 281 & 14.37 & 364.90 & 3621.80 & 0.3013 & 0.4650 & 0.6827 & 0.0182 \\
        \bottomrule
    \end{tabular}%
    }
\end{table*}

\begin{table}[!t]
    \centering
    \scriptsize
    \setlength{\tabcolsep}{2.5mm}
    \renewcommand{\arraystretch}{1.02}
    \captionsetup{skip=2pt}
    \caption{Transition-retrieval proxy under different summary budgets.}
    \label{tab:supp-summary-length-proxy}
    \begin{tabular}{@{}c r r r r r@{}}
        \toprule
        \textbf{Budget} & \textbf{Edges} & \textbf{Hit@1} & \textbf{Hit@3} & \textbf{Hit@5} & \textbf{MRR} \\
        \midrule
        10  & 1326 & 0.1026 & 0.4284 & 0.7934 & 0.3556 \\
        20  & 1326 & 0.1026 & 0.4615 & 0.7941 & 0.3605 \\
        30  & 1326 & 0.1026 & 0.4668 & 0.7753 & 0.3603 \\
        40  & 1326 & 0.1026 & 0.4465 & 0.7866 & 0.3559 \\
        60  & 1326 & 0.1026 & 0.4555 & 0.7443 & 0.3524 \\
        80  & 1326 & 0.1026 & 0.4955 & 0.7813 & 0.3653 \\
        100 & 1326 & 0.1026 & 0.4465 & 0.7677 & 0.3569 \\
        \bottomrule
    \end{tabular}
\end{table}

The results show that the word budget acts as an upper bound rather than a target length.
Even when the budget was increased to 60 or 100 words, the generated summaries usually remained around 15 words.
For activities, the average summary length increased from 12.13 words under the 30-word budget to 15.73 words under the 100-word budget; for widgets, it increased from 10.78 to 14.37 words.
Thus, larger budgets mainly changed which evidence was selected into the one-sentence summary, rather than forcing substantially longer outputs.

Very short summaries lost useful functional evidence.
Compared with the 30-word budget, the 10-word budget reduced activity evidence-term recall from 0.1443 to 0.1149 and action-object coverage from 0.1454 to 0.1053.
The drop was also clear for widgets: evidence-term recall decreased from 0.2446 to 0.1959, and action-object coverage decreased from 0.4116 to 0.3049.
The 20-word budget was closer to 30 words for activity summaries, but it was still slightly weaker for widget and action/object cues, which are important for modeling transition triggers.

Budgets above 30 words provided more room for local evidence terms, but did not consistently improve the transition proxy and could introduce more unsupported or redundant content.
For example, the 40-word budget increased some completeness metrics, but activity unsupported-term rate also increased from 0.7265 to 0.7363, and widget redundancy increased from 0.0076 to 0.0116.
At the transition-proxy level, Hit@3 decreased from 0.4668 at 30 words to 0.4465 at 40 words.
Longer budgets from 60 to 100 words also did not yield a monotonic advantage.
Token count and latency were not the main reason for choosing 30 words, because the fixed UI structure dominated the prompt length; from 30 to 100 words, average total tokens changed only slightly for both activities and widgets.

Therefore, the 30-word constraint should be interpreted as a controlled and pragmatic upper bound, not as a globally optimal length.
It is less restrictive than 10 or 20 words, which can omit transition-relevant cues, while still limiting the opportunity for unsupported details and redundant phrasing that can appear under looser budgets.
This provides a conservative balance between semantic completeness and noise for the saved summaries used by \builder.

\section{Additional Comparison Details: GCN vs.\ GIN Encoder}
\label{sec:supp-gcn-gin}

This section provides the encoder comparison removed from the main paper due to page limits.
Table~\ref{tab:rq11-encoder} presents the results of the two GNN encoders (GCN and GIN), under two thresholds ($@0.5$ and $@t^*$).
Based on the results, we have the following observations:
\begin{itemize}[leftmargin=2em, topsep=0pt, itemsep=0pt, parsep=0pt, partopsep=0pt]
    \item 
    \textbf{GIN consistently outperformed GCN across both thresholds.}
    Under $@0.5$, GIN attained a higher F1-score (0.9072) than GCN (0.8555), with gains in both precision and recall.
    Under $@t^*$, GIN reached 0.9158 F1-score, while GCN remained around 0.855.
    This trend was as expected, because GIN has greater discriminative ability than GCN in distinguishing graph structures~\cite{gin2019}.
    Unlike GCN, which relies on linear neighborhood smoothing~\cite{gcn2017}, GIN applies an MLP-based update after aggregation~\cite{gin2019}, which helps preserve discriminative signals and mitigates against over-smoothing~\cite{oversmoothing2018,oversmoothing2020}.
    Transitions in ATGs are typically hub-dominated: 
    (1) Many-to-one navigation means that many activities converge to a small set of high in-degree hub activities (e.g., ``back'' flows to \textttsplie{HomeActivity}); and 
    (2) one-to-many navigation means that a hub activity can have many outgoing transitions to diverse functional targets (e.g., a \textttsplie{HomeActivity} linking to multiple feature pages). 
    Over-smoothing in GCN could blur subtle differences among activities connected to the same hubs, reducing link-prediction separability.

    \item 
    \textbf{GIN yielded a higher validation-selected threshold than GCN.}
    The selected $t^*$ for GIN was higher ($\approx 0.4535$) than that for GCN ($\approx 0.3930$).
    This indicates that GIN could use a stricter decision threshold when converting predicted probabilities into binary transition predictions.
    Such a stricter threshold was consistent with better separation between the predicted probabilities of transition and non-transition edges:
    GIN could filter out more low-confidence edges (improving precision) while losing fewer true transitions (preserving recall) than GCN.
\end{itemize}

\rqsummary{GCN and GIN comparison}{Selecting an appropriate encoder for \builder\ (e.g., GIN) that effectively handles hub-dominated navigation graphs could improve the performance.
Accordingly, the prediction-threshold selection should be treated as part of the model configuration and tuned for the encoder (using a validation-selected $t^*$), rather than using a fixed value.}

\begin{table*}[!t]
    \centering
    \scriptsize
    \setlength{\tabcolsep}{1.78mm}
    \renewcommand{\arraystretch}{1.0}
    \captionsetup{skip=2pt}
    \caption{RQ1.1: Effect of encoder selection.}
    \label{tab:rq11-encoder}
    \begin{tabular}{@{}c c c c c c c c c c c@{}}
        \hline
        \multirowcell{2}{\textbf{ID}} &
        \multirowcell{2}{\textbf{Encoder}} & &
        \multicolumn{3}{c}{$\bm{@0.5}$} &
        \multicolumn{1}{c}{} &
        \multicolumn{4}{c}{$\mathit{\bm{@t^*}}$} \\
        \cline{4-6}\cline{8-11}
        & & & 
        \textbf{Precision} & \textbf{Recall} & \textbf{F1-score} &
        &
        \textbf{Precision} & \textbf{Recall} & \textbf{F1-score} & \textbf{$t^*$ Value} \\
        \hline
        \textbf{E1} & GCN & 
        & $0.8376_{\pm 0.0087}$ & $0.8744_{\pm 0.0189}$ & $0.8555_{\pm 0.0136}$
        &
        & $0.8271_{\pm 0.0025}$ & $0.8859_{\pm 0.0243}$ & $0.8551_{\pm 0.0120}$ & $0.3930_{\pm 0.0778}$ \\
        \textbf{E2} & GIN &
        & \hlo{\rwavo{$0.9057$}}$_{\pm 0.0121}$ 
        & \hlo{\rwavo{$0.9139$}}$_{\pm 0.0508}$ 
        & \hlo{\rwavo{$0.9072$}}$_{\pm 0.0321}$
        &
        & \hlb{\rwavb{$0.8995$}}$_{\pm 0.0131}$ 
        & \hlb{\rwavb{$0.9330$}}$_{\pm 0.0230}$ 
        & \hlb{\rwavb{$0.9158$}}$_{\pm 0.0177}$ 
        & $0.4535_{\pm 0.0423}$ \\
        \hline
    \end{tabular}
\end{table*}

\end{document}